\documentclass[sigconf,10pt]{acmart}

\usepackage[]{hyperref}
\usepackage{balance}
\usepackage{soul}

\def\Snospace~{\S}

\usepackage{xspace}
\newcommand{\ie}{\emph{i.e.,}\xspace}
\newcommand{\eg}{\emph{e.g.,}\xspace}
\newcommand{\etal}{\emph{et al.}\xspace}
\newcommand{\name}{FUSE\xspace}
\newcommand{\LLAMACPP}{\textit{llama.cpp}\xspace}

\usepackage[font=small]{caption}
\usepackage{subcaption}
\usepackage{listings}
\usepackage{algorithm2e}
\usepackage{enumitem}

\usepackage{multirow} % multirow
\usepackage{array,makecell} % thead

\newcommand{\takeaway}[1]{\noindent \textit{\textbf{Takeaways:} #1}}
\newcommand{\cut}[1]{{}}

\usepackage{tikz}
\newcommand{\tikzcircle}[2][black,fill=black]{\tikz[baseline=-0.5ex]\draw[#1,radius=#2] (0,0) circle ;}%

\AtBeginDocument{%
  }

\setcopyright{none}
\renewcommand\footnotetextcopyrightpermission[1]{} % removes footnote with conference information in first column
\begin{document}

\title{Dissecting the Impact of Mobile DVFS Governors on LLM Inference Performance and Energy Efficiency}

% \author{Submission \#679}

\author{Zongpu Zhang$^{1,2*}$,~Pranab Dash$^{2*}$,~Y. Charlie Hu$^{2}$,~Qiang Xu$^{2}$,\\Jian Li$^{1}$~and~Haibing Guan$^{1}$}
\affiliation{%
  \institution{$^{1}$Shanghai Jiao Tong University,~Shanghai,~China\\$^{2}$Purdue University,~West~Lafayette,~USA}
  \city{}
  \country{}
}

\thanks{* These authors contributed equally to this work.}
% \titlenote{Email: \{zhangz-z-p,~li-jian,~hbguan\}@sjtu.edu.cn,  \{dashp,~ychu,~xu1201\}@purdue.edu}

\renewcommand{\shortauthors}{}

\begin{abstract}
Large Language Models (LLMs) are increasingly being integrated into various
applications and services running on billions of mobile devices.
However, deploying LLMs on resource-limited mobile devices faces a
significant challenge due to their high demand for computation, memory, and
ultimately energy.
While current LLM frameworks for mobile
use three power-hungry components---CPU, GPU, and Memory---even
when running primarily-GPU LLM models,
optimized DVFS governors for CPU, GPU, and memory
featured in modern mobile devices
operate independently and are oblivious of each other.
Motivated by the above observation,
in this work, we first measure the energy-efficiency of a SOTA LLM framework consisting of various LLM models on
mobile phones
which showed the triplet mobile governors result in up to 40.4\%
longer prefilling and decoding latency compared to optimal combinations
of CPU, GPU, and memory frequencies with the same energy consumption
{for sampled prefill and decode lengths}.
Second, we conduct an in-depth measurement study to
uncover how 
the intricate interplay (or lack of)
among the mobile governors cause 
the above inefficiency
in LLM inference.
Finally,
based on these insights, we design FUSE --- a unified energy-aware
governor for optimizing the energy efficiency of
LLM inference on mobile devices.
Our evaluation {using a ShareGPT dataset} shows FUSE reduces the time-to-first-token
and time-per-output-token latencies by 7.0\%-16.9\% and 25.4\%-36.8\%
on average with the same energy-per-token
for various mobile LLM models.
\end{abstract}

\maketitle
\section{Introduction}
\label{sec:intro}

As Generative AI, powered by the LLM technology \cite{attention}, continues to evolve rapidly, it is
increasingly integrated into a wide range of societal
applications.
In particular, the technology is used to power personal services that
operate on billions of mobile devices, enabling users to access
advanced capabilities like personalized recommendations, intelligent
virtual assistants, and language translation at their fingertips.
In doing so, the proliferation of LLMs on mobile platforms is transforming how
individuals interact with technology, making it more intuitive and
accessible.

However, deploying LLMs on resource-constrained mobile devices
faces a significant challenge due to their high demand for
computation, memory, and energy. 
Mobile devices have limited processing capabilities, memory bandwidth,
and battery life, while LLM inference needs to perform
compute-intensive operations such as matrix-multiplications
as well as memory-intensive operations such
as accessing large K-V caches, both of which 
drive the CPU/GPU and memory to draw significant amount of power
and rapidly deplete the battery.
Addressing these challenges is critical to making LLM technology
more accessible and effective on mobile devices.

Several recent works (\eg~\cite{mobilellm,llmfirmware,minicpm-v,mobilequant},
%NPU
~\cite{mllm-npu}, and
%engines
~\cite{powerinfer2,llmcad,edgemoe,sti})
have focused on minimizing inference latency without considering
energy constraints. However, the energy consumption of on-device LLMs
remains exceptionally high, potentially limiting regular use by
everyday users.
For instance, a recent study \cite{meltingpoint}
reveals that battery could be depleted by inferencing only 490 to 590
prompts, using a mobile-tailored 4-bit quantized LLM on iPhone 14
Pro.

In this paper, we make two key observations about LLM inference serving
on mobile devices.
First, current LLM frameworks for mobile devices,
\eg llama.cpp~\cite{llamacpp},
use three power-hungry components: the CPU, GPU, and memory --
even when running primarily GPU-based LLM inference.
While it is expected that GPU-based LLM inference would extensively
use the GPU and memory intensively, the CPU also plays a significant
role. This is because the CPU remains actively engaged to support
OpenCL, the GPU programming framework commonly employed by
state-of-the-art mobile LLM frameworks.

Second, modern mobile operating systems (OSes)
feature optimized governors for
CPU, GPU, and memory that
perform
DVFS
to enhance energy-efficiency of respective components~\cite{web:eas:eas,web:aosp:gs201dtsigpu,web:arm:MIF2}.
But these governors are designed
to function independently
without coordination,
which can lead to suboptimal
energy efficiency across the system.
For example, when running a GPU-based LLM,
the CPU may scale down
its frequency to save CPU power without considering the high utilization
of the GPU, which relies on the CPU to quickly feed it the next operators,
potentially elongating the end-to-end inference latency and
increasing the energy per inference.

Motivated by the above observations, in this paper, we conduct to our
knowledge the first in-depth study of the interplay of mobile CPU,
GPU, and memory governors during LLM inference. Our study answers the
following questions.

{\bf (1) How well do the governors on modern mobile devices work for LLM inference workload?}
To study the impact of governors on LLM inference performance and
energy efficiency, we carefully design a benchmarking testbed
that can control the CPU/GPU/memory frequencies or use the default governors
of Android phones during each LLM inference run and automate a large number of inference runs
while measuring detailed inference performance (prefill vs. decode)
and power draw.
Using the testbed,
we search for the optimal CPU/GPU/memory frequencies that
minimize inference latency under the constraint of consuming no more energy
than using the default governors {for sampled prefill/decoding lengths}.
Our results
show that the  inference latency and energy efficiency
of the default governors can be far from optimal.
For example, fixing the components to operate at optimal frequency combinations
can either reduce the prefilling and decoding latency by up to 40.4\% or
reduce the energy consumption by up to 16.6\% compared to the default
governors.

{\bf (2) Why do the individually optimized governors
  together cause low performance and energy efficiency in LLM inference?}
To answer this question, we design a set of controlled experiments
to first isolate and gain insight into
the behavior and impact of individual
mobile governors (\eg by pinning the frequencies of other governors
as enabled by our benchmarking testbed),
and then examine the interplay when they act concurrently,
for LLM inference workload.
Our experiments provide several important insights:
(1) When the default governors work in isolation,
\ie trying to optimize the energy efficiency of respective components,
the GPU and CPU governors
tend to choose low frequencies that result in long inference latency
and low energy efficiency.
For example, the time-per-output-token
(TPOT) in decode under the GPU governor
which operates the GPU at 424.4 and 411.4 MHz for
TinyLlama-1.1B and StableLM-Zephyr-3B models
could be reduced by 41.0\% and
34.6\% for the two models
without increasing energy consumption, if it had chosen higher GPU frequencies (848 and 762 MHz).
Similarly, the TPOT in decode for the two models under
the EAS governor,
which chose overly low CPU frequencies of 1130.8 and 1038.8 MHz, respectively,
could be reduced by 13.2\% and 13.4\%, if EAS had chosen higher CPU frequencies of 2252 and 2401 MHz,
for the two models, respectively.
(2) When the GPU and CPU governors act concurrently, they can
antagonistically trigger a "downward spiral" effect where they
drive each other to cascadingly lower the CPU/GPU frequencies,
which explains the long inference latency and low energy efficiency
observed in {\bf (1)}.
We further uncover the root cause for the antagonistic effect, which
stems from each governor independently scaling its frequency in an
attempt to achieve its designated utilization target.

{\bf (3) How to design a unified energy-aware governor to optimize the
  energy efficiency of all the three power-hungry components for LLM
  inference?}
Avoiding such antagonistic effect between independently
acting governors uncovered from our root cause analysis above
requires a holistic energy-efficient governor for
managing all three components.
To this end, we present \name, a unified Energy-aware
Governor for optimizing the energy efficiency of LLM inference on mobile devices.
\name operates in two stages.
At the offline stage,
it performs efficient profiling-based search
for the optimal frequency combination for the three
power-hungry components that
minimizes the inference latency given an energy budget
or minimizes the energy consumption given an inference latency
target, \ie TTFT for prefill and TPOT for decode.
During runtime, for each inference request,
it looks up and pins down the components at the optimal frequency combination.
We prototype \name on Android Pixel 7 and Pixel 7 pro phones and
evaluate it using the ShareGPT dataset.  Our results show that \name reduces the TTFT and TPOT by 7.0\%-16.9\% and 25.4\%-36.8\%
on average
for various mobile LLM models, compared with the default
mobile governors.

We have released \name as an extension to the \LLAMACPP framework
to facilitate further research on energy-efficient LLM inference.

\section{Background and Motivation}
\label{sec:background_and_motivation}

\subsection{Computational Characteristics of LLM Inference}

The inference of most popular LLM models, \eg the
GPT~\cite{brown2020languagemodelsfewshotlearners} and
Llama~\cite{touvron2023llamaopenefficientfoundation} series, is done in an
autoregressive manner, which consists of two stages: the {\em prefill} stage, where
the user prompt is processed to generate the first token of the response, and 
the {\em decode} stage, where
subsequent tokens are generated one by one until a special end-of-sequence token is reached.
Both stages run the same LLM model, which
consists of multiple (32 for Llama-2~7B~\cite{touvron2023llama2openfoundation})
Transformer blocks, and each Transformer block is in turn composed of an
attention component and an MLP component~\cite{attention}. The LLM model only runs once during
the prefill stage, where tokens from the user prompt are processed in a batch,
which is very compute-intensive. On the other hand, during the decode stage,
the model runs once for each output token. However, with the widely-used
KV-cache optimization, only the last token needs to be processed by the model
in order to generate the next token. Thus, the model essentially runs with
batch size 1 and is less compute-intensive. The different computational
characteristics of the two stages 
often require different optimization configurations~\cite{10609649,298687}
that also entail different energy characteristics (as
we will see in \autoref{sec:energy-efficiency}).

\subsection{LLM Inferencing Uses Multiple Hardware Components}
\label{subsec:inference}

Current LLM frameworks for mobile~\cite{llamacpp,mlc-llm,alibaba2020mnn,mllm-npu},
for example, \LLAMACPP~\cite{llamacpp},
use all three power-hungry components, CPU, GPU, and memory,
and this is true even when running primarily GPU-based  LLM models.
LLM models are generally more efficient running on mobile GPUs than on mobile CPUs~\cite{meltingpoint}.
Such GPU-based LLM models typically use OpenCL, the prominent programming framework for mobile GPUs.
OpenCL provides an asynchronous abstraction where the sequence of kernels is
enqueued into a command queue, and the OpenCL client only needs to
wait for the GPU to finish executing the enqueued kernels without
synchronizing with the GPU in the middle. However, compared to desktop
GPUs, mobile GPUs have limited hardware support for queue
management. For example, the ARM Mali GPUs only have a shallow queue
with 2 outstanding entries at maximum~\cite{10.1145/3503222.3507754}. Because
of this, the OpenCL
library (running on the CPU) still needs to manage a much deeper
command queue and feed new kernels to the GPU as those queued on the
GPU finish. As a result, the CPU is constantly involved over the entire
duration of LLM inference, even though most computation is offloaded to
the mobile GPU.

\subsection{Mobile DVFS Governors}
\label{subse:governors}

\begin{figure}[tp]
\captionof{table}{Available frequencies in Pixel 7 and Pixel 7 Pro}
\vspace{-3mm}
\begin{minipage}[c]{\linewidth}
\centering
\resizebox{\textwidth}{!}{
\begin{tabular}{ *{2}{rl} }
    \toprule
    \bf Governor & \bf Available Frequencies (MHz) \\
    \midrule
    CPU & 500, 851, 984, 1106, 1277, 1426, 1582, 1745, 1826, 2048, 2188, 2252, \\
    & 2401, 2507, 2630, 2704, 2802, 2850\\
    \midrule
    GPU & 151, 202, 251, 302, 351, 400, 471, 510, 572, 701, 762, 848 \\
    \midrule
    Memory & 421, 546, 676, 845, 1014, 1352, 1539, 1716, 2028, 2288, 2535, 2730, 3172 \\  
    \bottomrule
\end{tabular}}
\end{minipage}
\label{tab:profile-freq}
\vspace{-4mm}
\end{figure}

Mobile operating systems such as Android employ DVFS governors
which perform Dynamic Voltage and Frequency Scaling (DVFS)
to optimize the energy efficiency of
power-hungry hardware components such as the GPU, CPU, and memory.
Intuitively, the energy efficiency of a component is a function of its
power draw and runtime in completing a given workload,
$e = energy / load = power*runtime/load$.
Since in general increasing operating frequency 
reduces runtime but increases power,
and vice versa, the governor aims to choose a target
frequency that balances the two to achieve high energy efficiency.
In practice,
since neither power nor runtime can be easily measured,
a practical governor design is to adjust the frequency
based on observed systems statistics such as utilization of the component,
\ie trying to keep the 
component running around a target utilization, by
increasing (decreasing) the operating frequency to the next frequency
(based on a pre-defined utilization-to-frequency lookup table)
if the hardware utilization exceeds (drops below)
some threshold.

Off-the-shelf governors used in mobile operating systems such as Android
are designed and optimized for the respective hardware components in isolation;
at runtime, they function independently of each other.

\begin{figure}[tp]
    \centering
    \begin{subfigure}[]{0.40\textwidth}
    \centering
    \includegraphics[width=0.80\linewidth]{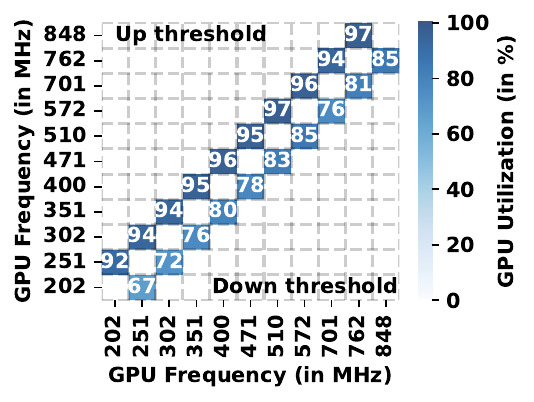}
    \label{fig:util-freq-gpu}
    \end{subfigure}
    \if 0
    \centering
    \begin{subfigure}[]{0.48\textwidth}\includegraphics[width=0.98\linewidth]{Figures/Sec2-Background/Sec2-MaiQuickstepGPUGovernor.pdf}
      \caption{GPU frequency adjustment during \LLAMACPP inference on Pixel 7}
    \label{fig:Timeline_gpu_governor}
    \end{subfigure}
    \fi
    \vspace{-0.2in}
    \caption{Target utilization range per GPU frequency}
    \label{fig:Gpu_governor}
    \vspace{-5mm}
\end{figure}

\textbf{GPU governor.}  
The Quickstep~\cite{web:aosp:gs201dtsigpu} governor
is used to manage the GPU frequency in recent generations of phones
such as the Google Pixel family from Pixel 6 to the latest Pixel 9.
\cut{For our experiments,
we used Pixel 7 and Pixel 7 Pro which have ARM Mali-G710 MP7 GPU and this governor.}
It determines the target frequency based
on a predefined table of minimum and maximum utilization for each GPU
frequency which is provided by the manufacturers.
Fig.~\ref{fig:Gpu_governor} %~\ref{fig:util-freq-gpu}
depicts the threshold limits (found from \texttt{dvfs\_table})---if the utilization falls below
the minimum utilization, the frequency is scaled down and if it
exceeds the maximum utilization, the frequency is scaled up.

\textbf{Memory governor.}  The memory interface
(MIF)~\cite{web:arm:MIF1,web:arm:MIF2,web:arm:MIF3} connects the main
memory to all the system components such as the CPU and GPU.  The number
of read and write transactions with the main memory is affected by MIF
frequency, \eg if the MIF frequency is too low then the CPU will incur
higher data latency and will start stalling.  As running tasks 
switch between compute and memory phases,
there will be bursts of memory transactions interleaved with idle or lightly loaded periods.
The main objective of MIF is to deliver data with minimum latency during
such burst periods.
To this end, MIF uses an interactive governor which
increases memory frequency to the peak when it observes high memory bus
utilization,
and steps down the frequency as the utilization drops,
by caculating a target frequency based on
a formula and a predefined
factor~\cite{web:aosp:governorsimpleinteractive}
provided by the manufacturer,
and setting the MIF frequency to one of the 13 frequencies from
Table~\ref{tab:profile-freq} closest to the target frequency. The
above process is repeated every 20
ms~\cite{web:interactive:patch,web:arm:governors,banerjee2018characterization1,banerjee2018characterization2}.

\textbf{CPU governor.}
Android smartphones employ Energy-aware
scheduling (EAS)~\cite{web:eas:eas} for CPU power management, which
encompasses both task placement and DVFS control.  Deciding the right
frequency for a given task is done in two steps.
(1)
First, EAS determines the \textit{load} of the task.
A straightforward estimation of a task's load is its CPU utilization.
However, such an estimation is not frequency-invariant, as the CPU
utilization is usually higher under lower frequency. To this end, EAS
applies different scaling factors to the CPU utilizations measured
under different frequencies, such that the scaled loads for the same task
stay roughly the same.
The current load of a task is estimated from its
historical loads sampled every millisecond, with earlier samples
exponentially decayed.
(2)
Second, EAS is provided a per-cluster load-to-frequency lookup table that is used to
find the lowest frequency that can satisfy the estimated task load.
The task load continues to decay when the task is waiting for I/O or other
resources, \eg GPU~\cite{Corbet13}. Thus, EAS may choose lower CPU frequencies
for GPU-heavy tasks like LLM inference due to the low task load.

\subsection{Research Questions}

The above discussion highlights two observations:
(1) Current LLM frameworks for mobile
use three power-hungry components---CPU, GPU, and Memory---even
when running primarily-GPU-based LLM inference;
(2) DVFS governors employed in mobile OSes such as Android
are designed
to optimize the energy efficiency
for individual components (CPU, GPU, and memory).
As such, they
are oblivious to each other's dynamic adjustments.
Such lack of coordination of different governors
can potentially lead to suboptimal energy efficiency across the system,
and motivates the following research questions:
\begin{itemize}[wide, labelwidth=!, labelindent=0pt]
\item How well do the governors on modern mobile devices work for LLM inference workload? (\S\ref{sec:energy-efficiency})
\item How does the intricate interplay (or lack of) among governors cause the energy inefficiency in LLM inference? (\S\ref{sec:interplay})
\item How to design a unified energy-aware governor to optimize the overall energy efficiency of
  all three hardware components involved in LLM inference? (\S\ref{sec:design})
\end{itemize}

To answer these questions,
we start with an in-depth measurement study 
by comparing the energy drain and performance
of LLM inference under the default DVFS governors
and under controlled operating frequencies of all
three components (CPU, GPU, and memory).

\section{Methodology}
\label{sec:methodology}

To enable detailed performance and power draw
measurements for studying the effectiveness of the default mobile
governors in on-device LLM inference, we carefully engineered
a benchmarking testbed. 

\textbf{Platform and power measurement.}  Our hardware platform consists of
several mid-tier smartphones, Google Pixel 7 and Pixel 7 Pro,
with Google Tensor G2 CPU and Mali-G710 MP7 GPU,
and 
running stock Android 13.
Pixel 7 has 8GB DRAM, while Pixel 7 Pro has 12GB DRAM.
The phones are rooted and opened; their batteries are bypassed and the
phones are powered by Monsoon power monitors~\cite{web:monsoon} which report fine-grained
power draw every 0.2 ms.
Since adb~\cite{web:adb} is unavailable during the experiment, we implement a profiling
daemon on the phone to automatically execute benchmark scripts. The
screen is turned off during profiling so it does not draw power.

\textbf{Governors.}  The Pixel 7 and Pixel 7 Pro phones use a set of
contemporary governors by default: \texttt{sched-pixel} EAS for CPU,
\texttt{quickstep} governor for GPU,
and \texttt{interactive} governor for memory interface.
To compare the performance/energy drain of LLM inference under
these governors with under alternative
configurations of CPU/GPU/memory frequencies,
we need a way to pin down the
CPU/GPU/memory to a given frequency combination.
We achieve this by leveraging the default governors.
Specicially,
since each component's
governor operates between a minimum and a maximum frequency of the component,
we can pin down 
the component (\eg GPU) to a fixed frequency (\eg $f_{GPU}$)
by writing it
into
the target minimum (\texttt{scaling\_min\_freq} or \texttt{min\_freq}) and maximum (\texttt{scaling\_max\_freq} or \texttt{max\_freq}) frequencies of the  governor.
We denote this setting as \texttt{Pin}.

\textbf{LLM framework and models.}
Our testbed focuses on the popular \LLAMACPP~\cite{llamacpp} framework, which is a C++ library to perform efficient, cross-platform inference of LLMs with a focus on optimizing tensor operations for performance.
We select a set of LLM models
that are widely used in research studies and in real mobile applications, \ie TinyLlama 1.1B \cite{tinyllama}, StableLM-Zephyr 3B, and Llama-2 7B \cite{touvron2023llama2openfoundation}.
Due to memory constraints, all Llama-2 \cite{touvron2023llama2openfoundation} experiments run on Pixel 7 Pro,
while other experiments run on Pixel 7.
Prior works such as \cite{meltingpoint} have reported that offloading LLMs to mobile GPU can achieve higher energy efficiency than running on CPU, so
we enable GPU inferencing in the \LLAMACPP framework with
OpenCL support by linking it to the CLBlast library~\cite{clblast}.
When performing inference, 
models are launched on the performance core (ARM Cortex-X1) while other
profiling processes are pinned to the LITTLE core (ARM Cortex-A55).

\textbf{Metrics.}  To quantify inference {\em performance}, we report
time-to-first-token (TTFT) for the prefill stage and
time-per-output-token (TPOT) for the decode stage.
We consider the typical mobile LLM usage scenario where the phone user
initiates a new inference request after the previous request has
returned, \ie the input queuing time in the prefill stage is not
considered in TTFT.
The end-to-end latency (E2E) is also
reported in some of the
experiments as appropriate. \cut{estimated as $TTFT + (N_{d}-1)
* TPOT$, where $N_d$ denotes the decode length (number of tokens).}
To quantify inference {\em energy-efficiency}, we report energy-per-token, \ie how
much energy (of the device) is consumed per prefill or decode token,
calculated as $P * TTFT / N_p$ and $P * TPOT$ for prefill
and decode stages, respectively,
where $P$ denotes the average power
consumption during the prefill or decode stage,
and $N_p$ denotes the prefill prompt length (number of tokens).

\section{Optimality of Mobile DVFS Governors}
\label{sec:energy-efficiency}

\begin{figure}[!t]
\begin{minipage}[c]{.49\linewidth}
\centering
\includegraphics[width=\linewidth, trim=0 0 0 0]{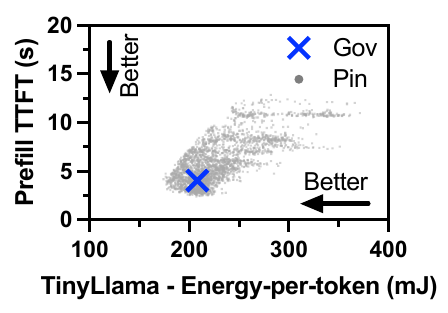}
% \subcaption{TinyLlama Prefill}
\end{minipage}\hfill
\begin{minipage}[c]{.5\linewidth}
\centering
\includegraphics[width=\linewidth, trim=0 0 0 0]{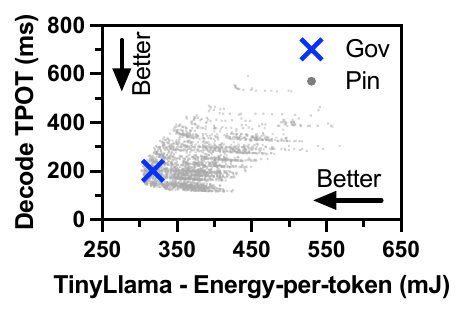}
% \subcaption{TinyLlama Decode}
\end{minipage}
% newline
\begin{minipage}[c]{.49\linewidth}
\centering
\includegraphics[width=\linewidth, trim=0 0 0 0]{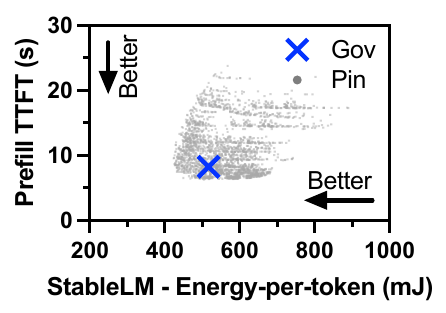}
% \subcaption{StableLM Prefill}
\end{minipage}\hfill
\begin{minipage}[c]{.5\linewidth}
\centering
\includegraphics[width=\linewidth, trim=0 0 0 0]{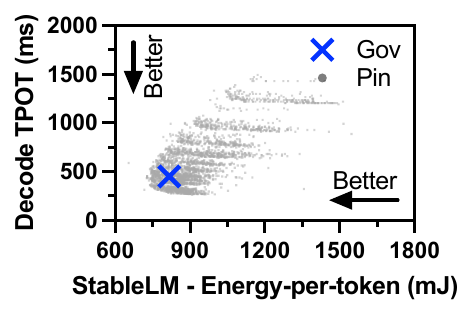}
% \subcaption{StableLM Decode}
\end{minipage}
\newline
\vspace{-2mm}
\caption{Comparison of inference latency and energy drain under default governors (Gov) and
  under different frequency combinations (Pin). 
  }
\label{fig:finding-baseline}
\vspace{-0.2in}
\end{figure}

We begin our study by measuring the latency and energy-efficiency of LLM
inference under the default mobile DVFS governors (denoted as \texttt{Gov}),
and
under all CPU, GPU, and memory frequency combinations (denoted as \texttt{Pin}).
Fig.~\ref{fig:finding-baseline} shows the latency and energy
consumption of every frequency combination (grey dots) along with
default governors ("$\times$" markers)
for LLM inference with 32 prefill tokens and 32 decode tokens using two
different models.
There are $18*12*13=2808$ frequency
combinations in total, as listed in Table \ref{tab:profile-freq}.
We found that among all \texttt{Pin} frequency combinations, many
are able to achieve better inference latency \textit{and} lower
energy consumption at the same time (the lower left region of each figure)
compared to the governors. With the same energy consumption as
\texttt{Gov}, prefill TTFT and decode TPOT can be reduced by up to
40.4\% and 31.8\% for TinyLlama, and up to 23.0\% and 37.1\% for
StableLM, respectively. On the other hand, with the same TTFT or TPOT
as \texttt{Gov}, energy-per-token can be reduced by up to 14.9\% and
5.0\% in prefill and decode for TinyLlama; and up to 16.6\% and 12.3\%
for StableLM, respectively.

\begin{figure}[!t]
\includegraphics[width=\linewidth, trim=0 10 0 0]{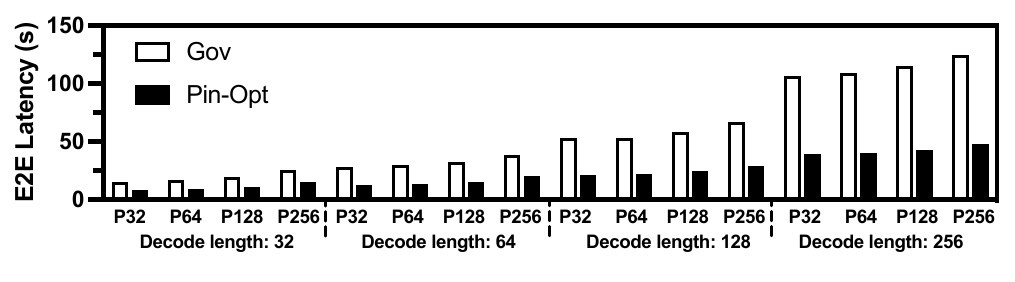}
\caption{End-to-end inference latency of TinyLlama on mobile with
  default governors (Gov) compared with running at the optimal frequency combination
  (Pin-Opt)
  that consumes the same amount of energy as with default
  governors.
  \texttt{P32}, \texttt{P64},
  \texttt{P128}, \texttt{P256} refer to prefill length of 32, 64, 128,
  256 tokens.}
\label{fig:finding-baseline-e2e}
\vspace{-4mm}
\end{figure}

To see the trend with different sequence lengths, we conduct additional
experiments with prefill and decode lengths being 32, 64, 128, or 256 tokens
(in total 16 combinations).
Fig.~\ref{fig:finding-baseline-e2e}
compares the end-to-end inference latency under \texttt{Gov} against the fastest
\texttt{Pin} latency combination with the same energy consumption for the
TinyLlama model, denoted as \texttt{Pin-Opt}.
We see \texttt{Pin-Opt} consistently achieves shorter end-to-end latency.
For instance, with 128 prefill tokens and 256 decode tokens,
\texttt{Pin} reduces end-to-end latency from 115.15 seconds to 42.50
seconds (63\% reduction). On average, end-to-end latency under
\texttt{Pin} is 54.9\% lower than
under \texttt{Gov} across the 16 combinations.

\begin{figure*}[!t]
  \begin{minipage}[c]{.32\linewidth}
\centering
\includegraphics[width=1.1\linewidth, trim=0 15 0 0]{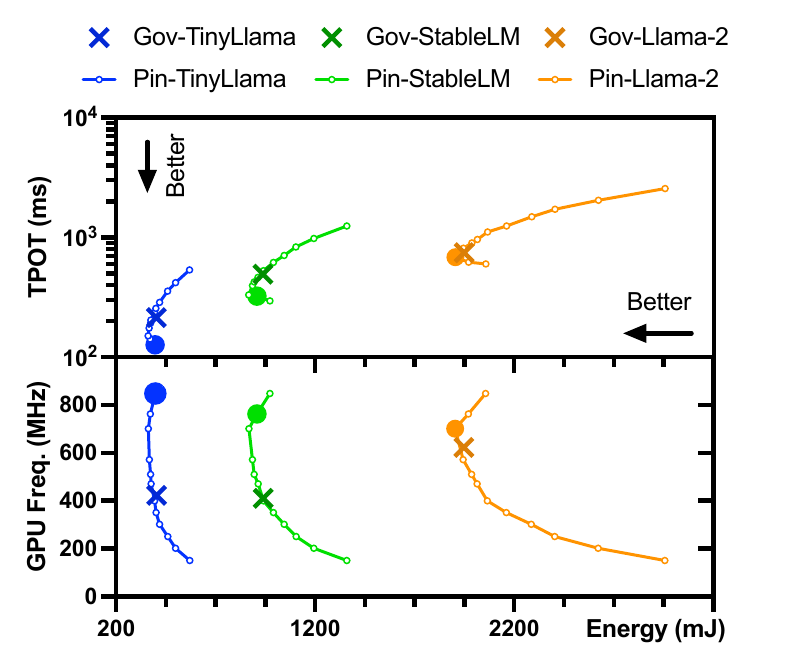}
\subcaption{Various Models}
\label{fig:finding-gpu-decode-a}
\end{minipage}\hfill
\begin{minipage}[c]{.32\linewidth}
\centering
\includegraphics[width=1.1\linewidth, trim=0 15 0 0]{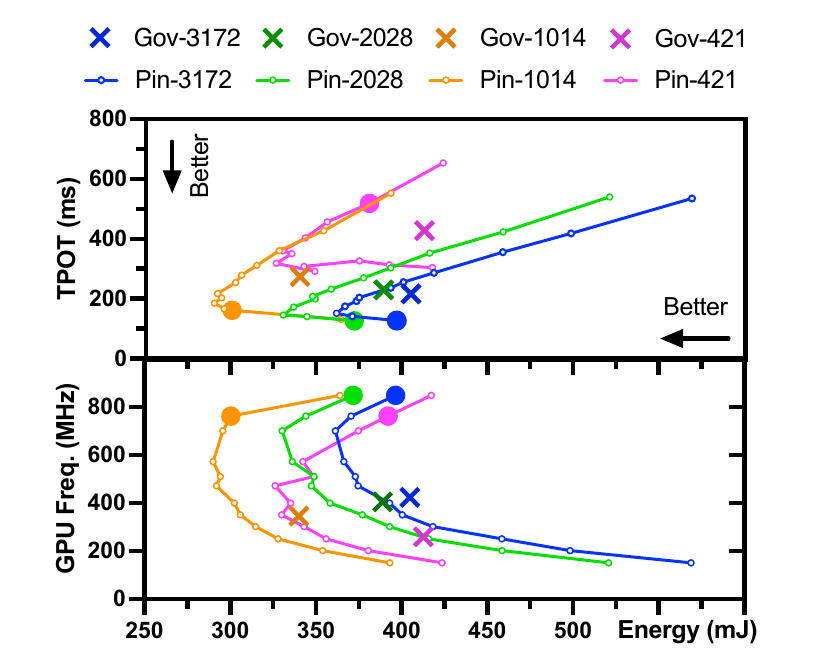}
\subcaption{Varying pinned $f_{MEM}$}
\label{fig:finding-gpu-decode-b}
\end{minipage}\hfill
\begin{minipage}[c]{.32\linewidth}
\centering
\includegraphics[width=1.1\linewidth, trim=0 15 0 0]{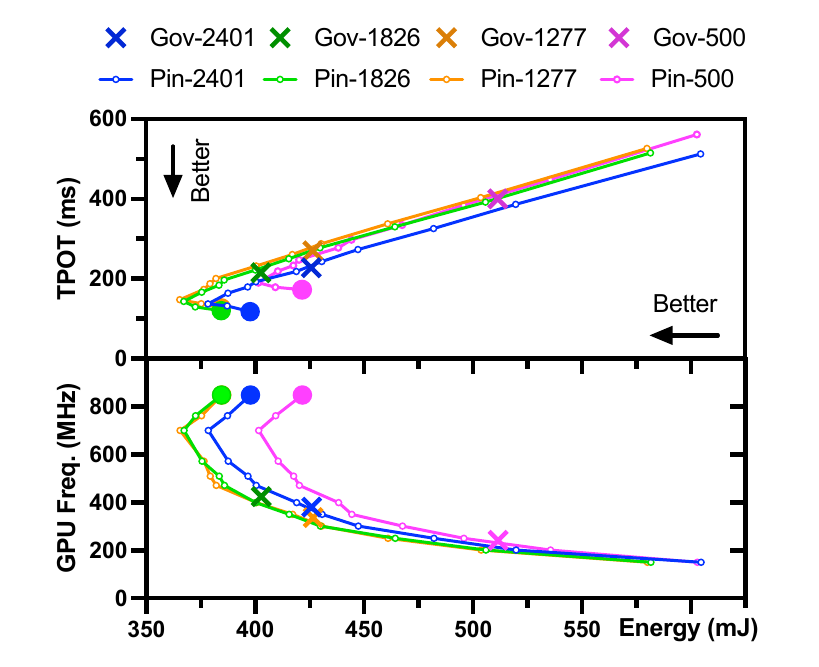}
\subcaption{Varying pinned $f_{CPU}$}
\label{fig:finding-gpu-decode-c}
\end{minipage}
\vspace{-0.15in}
\caption{Decode latency and energy-per-token of the GPU governor (\texttt{Gov})
  compared with pinning GPU at each available frequency (\texttt{Pin}). We
  set $f_{CPU}$=1826 MHz in (a) and (b), and $f_{MEM}$=3172 MHz
  in (a) and (c).
  Plots in (b) and (c) are for TinyLlama. {The lowest-latency frequency combinations with the same energy drain as the GPU governor, \texttt{Pin-Opt}, is marked with }\tikzcircle{2pt}.}
\label{fig:finding-gpu-decode}
\vspace{-3mm}
\end{figure*}

In summary, the above results show that the
energy efficiency and inference latency under
the default governors are far from optimal.

\section{\mbox{\hspace{-0.1in}}Understanding Impact of DVFS Governors}
\label{sec:interplay}

\begin{figure*}[!t]
  \hspace{-0.1in}
  \begin{minipage}[c]{.32\linewidth}
\centering
\includegraphics[width=1.1\linewidth, trim=0 15 0 0]{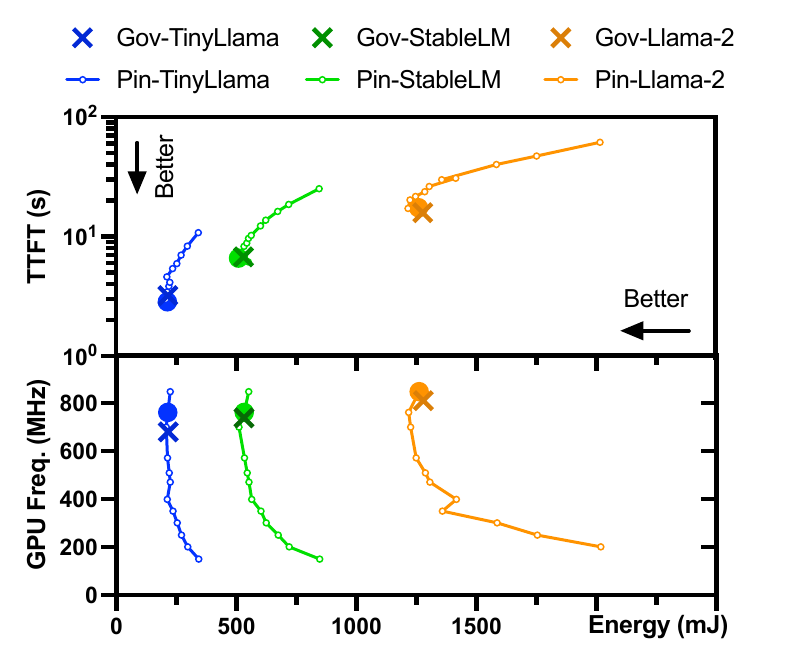}
\subcaption{Various Models}
\label{fig:finding-gpu-prefill-a}
\end{minipage}\hfill
\begin{minipage}[c]{.32\linewidth}
\centering
\includegraphics[width=1.1\linewidth, trim=0 15 0 0]{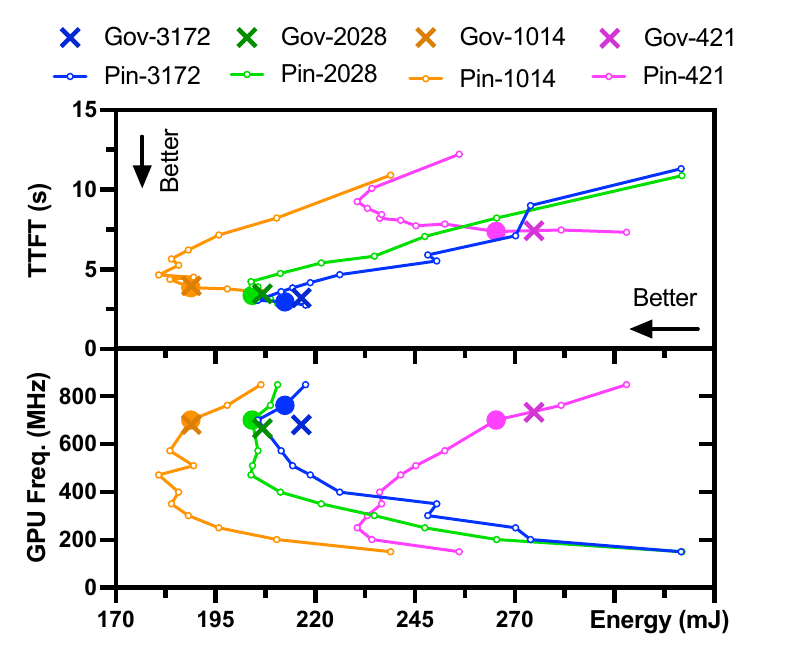}
\subcaption{Varying pinned $f_{MEM}$}
\label{fig:finding-gpu-prefill-b}
\end{minipage}\hfill
\begin{minipage}[c]{.32\linewidth}
\centering
\includegraphics[width=1.1\linewidth, trim=0 15 0 0]{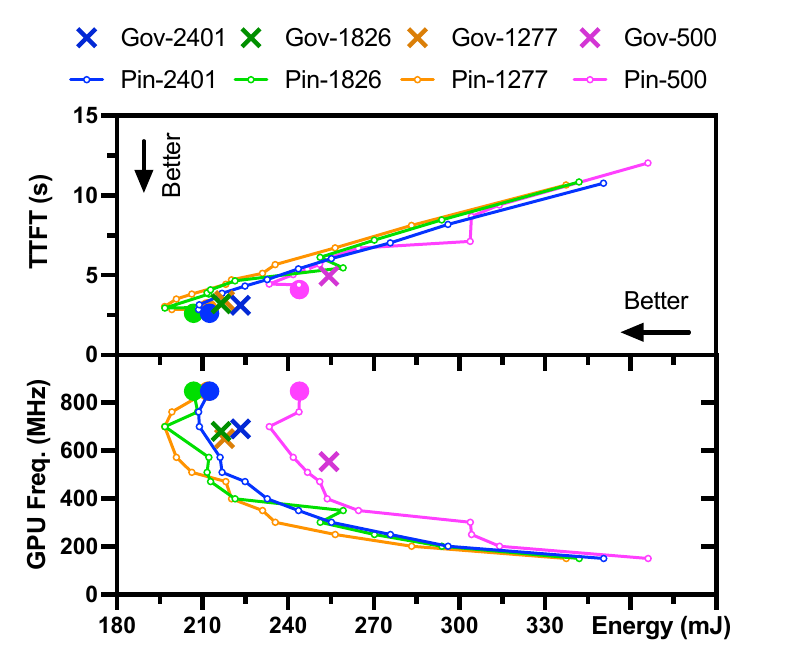}
\subcaption{Varying pinned $f_{CPU}$}
\label{fig:finding-gpu-prefill-c}
\end{minipage}
\vspace{-0.15in}
\caption{Prefill TTFT and energy-per-token of the GPU governor (\texttt{Gov}) 
  compared with pinning the GPU at each available frequency (\texttt{Pin}). We set $f_{CPU}$=1826 MHz in (a) and (b), and $f_{MEM}$=3172 MHz in (a) and (c). Plots in (b) and (c) are for TinyLlama. {The lowest-latency frequency combinations with the same energy drain as the GPU governor, \texttt{Pin-Opt}, is marked with }\tikzcircle{2pt}.
}
\label{fig:finding-gpu-prefill}
\vspace{-4mm}
\end{figure*}

To understand the interplay (or lack thereof) among DVFS governors during LLM inference
and its impact on inference performance and energy efficiency,
we design controlled experiments to first isolate the behavior
of individual governors and then examine
their cascading effect on one another.

\subsection{GPU Governor is only GPU-Energy Aware}
\label{subsec:gpu-gov}

To isolate the impact of GPU governor on LLM inference from 
other governors, we pin CPU and memory frequencies and compare LLM
inference under the default GPU governor vs. when the GPU frequency is
pinned to each available frequency using {\tt Pin}.
Due to page limit, we show results for prefill and decode
lengths fixed to 32 tokens; the results for other prefill/decode
length combinations are similar.

Since the actual GPU frequency under the default GPU governor can vary
during an LLM inference, for intuitive comparison
of results under the governor vs. under individual pinned GPU frequencies,
we report a single, {\em effective frequency},
calculated as the weighted average of each observed frequency
during inference, for each inference run under the governor;
the weight for each frequency is the percentage of time the governor stays
at that particular frequency.

\textbf{Decode.}
In the first set of experiments, we pin the CPU and memory frequencies
to fixed values.
Fig.~\ref{fig:finding-gpu-decode}(a) upper half shows
the decode TPOT vs.~energy consumption per token
under the GPU governor compared with
when the GPU is pinned at each available frequency
for various LLM models.
Overall, the GPU governor fails to achieve either latency or energy optimality.
{
For instance, it achieves 215.1 ms TPOT at 402.7 mJ per token for TinyLlama, while
pinning the GPU at 848 MHz achieves 41.0\% lower latency (126.9 ms) with similar energy drain (396.5 mJ).
For StableLM, it achieves 495.0 ms TPOT at 937.5 mJ per token, while pinning the GPU at 762 MHz achieves 34.6\% lower latency (323.6 ms) with similar energy drain (907.2 mJ).
Alternatively, \texttt{Pin} achieves 7.0\% and 7.6\% lower energy with similar latency for the two models respectively.
}

To understand why the GPU governor cannot achieve the lowest TPOT,
we plot the corresponding GPU frequencies for all the inference runs
in Fig.~\ref{fig:finding-gpu-decode}(a) lower half.
We see that the GPU governor runs at overly low
frequencies for TinyLlama and StableLM; the effective
GPU frequencies for the two models are 424.4 MHz and 411.0 MHz,
respectively.

The TPOT under the GPU governor for larger models is closer to the
optimal.  This is because decoding a larger model is more
compute-intensive (with a decode length of 32 tokens, TPOT is
dominated by MLP layers in LLM) and drives the GPU to run at higher
utilization, allowing the governor to boost the GPU frequency to be
sufficiently high (\S\ref{subse:governors}). For instance, for Llama-2, the effective frequency
under the GPU governor is 624.1 MHz and the TPOT is only 8.7\% higher
than the optimal GPU frequency constrained by the same energy
consumption.

\begin{figure*}[!t]
  \hspace{-0.1in}
  \begin{minipage}[c]{.32\linewidth}
\centering
\includegraphics[width=1.1\linewidth, trim=0 15 0 0]{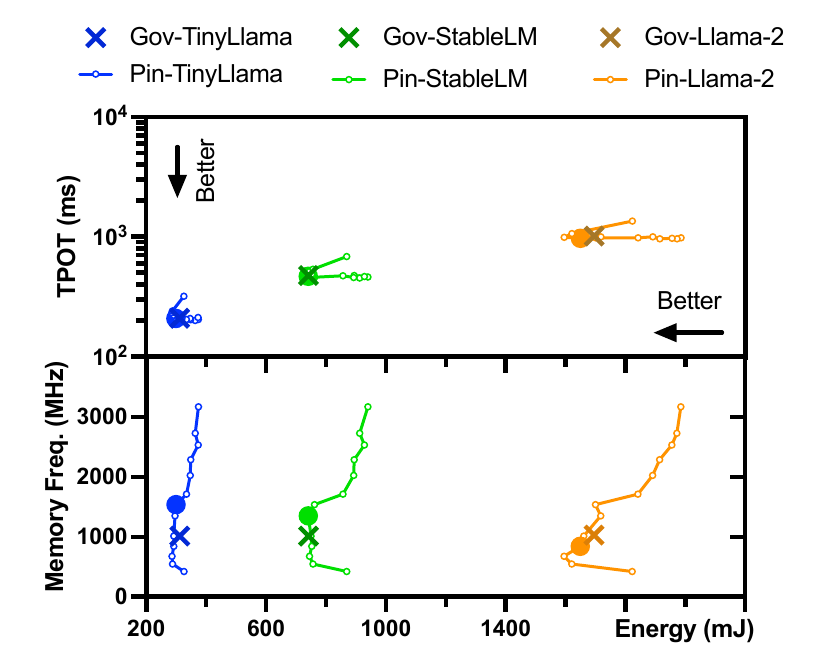}
\subcaption{Various Models}
\label{fig:finding-mem-decode-a}
\end{minipage}\hfill
\begin{minipage}[c]{.32\linewidth}
\centering
\includegraphics[width=1.1\linewidth, trim=0 15 0 0]{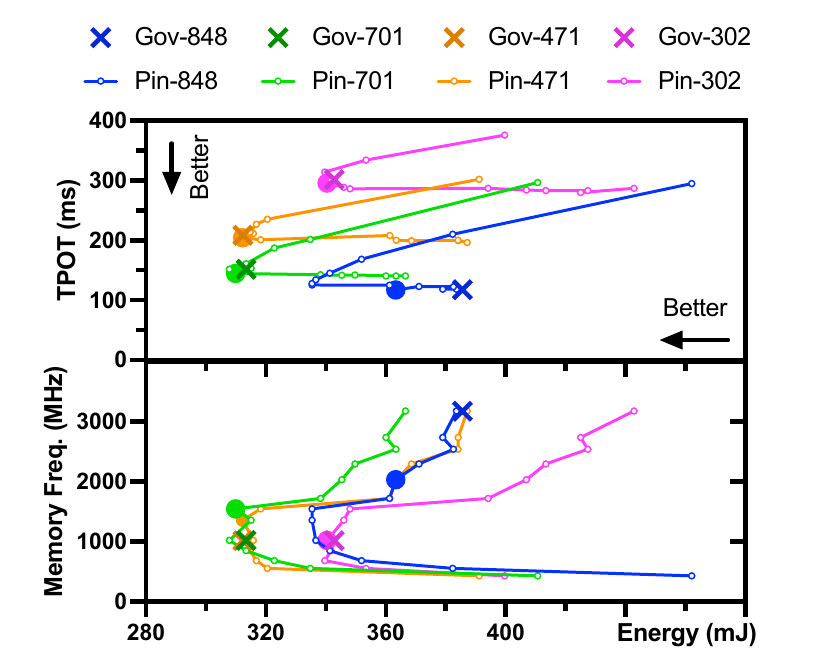}
\subcaption{Varying pinned $f_{GPU}$}
\label{fig:finding-mem-decode-b}
\end{minipage}\hfill
\begin{minipage}[c]{.32\linewidth}
\centering
\includegraphics[width=1.1\linewidth, trim=0 15 0 0]{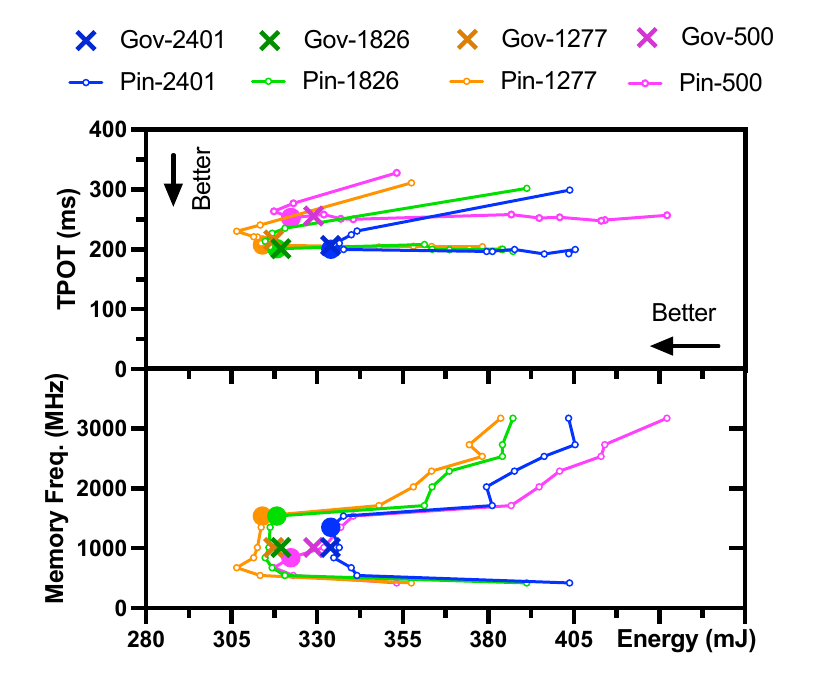}
\subcaption{Varying pinned $f_{CPU}$}
\label{fig:finding-mem-decode-c}
\end{minipage}
\vspace{-0.15in}
\caption{Decode latency and energy drain of the memory governor (\texttt{Gov})
 compared with pinning memory at each available
  frequency (\texttt{Pin}). GPU and CPU frequencies are pinned at 471 MHz and
  1826 MHz.
  Plots in (b) and (c) are for TinyLlama.  {The lowest-latency frequency combinations with the same energy drain as the memory governor, \texttt{Pin-Opt}, is marked with }\tikzcircle{2pt}.}
\label{fig:finding-mem-decode}
\vspace{-4mm}
\end{figure*}

\begin{figure}[!t]
\centering
\begin{minipage}[c]{.5\linewidth}
\centering
\includegraphics[width=\linewidth, trim=0 10 0 0]{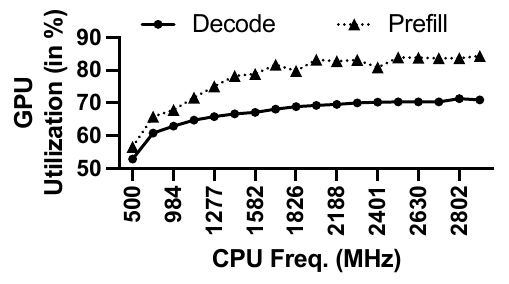}
\end{minipage}\hfill
\begin{minipage}[c]{.5\linewidth}
\centering
\includegraphics[width=\linewidth, trim=0 10 0 0]{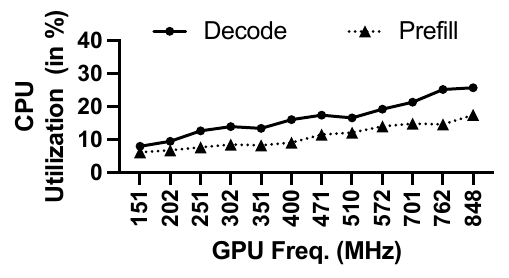}
\end{minipage}
\caption{GPU utilization with pinning CPU at each available frequency and $f_{GPU}$=701 MHz (left), and CPU utilization with pinning the GPU at each available frequency and $f_{CPU}$=2188 MHz (right). Plots are for TinyLlama and $f_{MEM}$=3172 MHz.}
\label{fig:util-freq}
\vspace{-4mm}
\end{figure}

In the second set of experiments, we focus on the TinyLlama model
and repeat the above experiments while
varying the set of frequencies that the memory is pinned to.
Fig.~\ref{fig:finding-gpu-decode}(b) shows the results.
We see when varying the pinned memory frequency,
the GPU governor consistently chooses overly low frequencies
which result in high TPOT and energy-per-token.
For instance, when the memory is pinned to a medium 
frequency of 1014 MHz, the effective GPU frequency under the GPU
governor is 346.5 MHz,
which achieves 273.9 ms TPOT and 340.0 mJ energy-per-token,
while pinning the GPU at 762 MHz results in 41.0\% lower TPOT (161.6 ms) and a similar
energy consumption (300.4 mJ).
Further, lower memory frequencies appear to lead the GPU governor to
reduce GPU frequency to maintain sufficient GPU utilization, \eg the
effective GPU frequencies under the GPU governor given 3172, 2028,
1014, and 421 MHz pinned memory frequencies are 424.4, 406.0,
346.5, and 259.7 MHz, respectively.

In the third set of experiments,
we fix memory frequency and repeat the above experiments
under varying pinned CPU frequency.
As shown in Fig.~\ref{fig:finding-gpu-decode}(c), medium to
high CPU frequencies have limited impact on the GPU governor. Overly low
CPU frequency, such as 500 MHz, causes the GPU governor to frequently
choose lower GPU frequencies which results in high TPOT and
energy-per-token compared to the optimal configuration (high GPU frequency).

The reason that the GPU governor tends to choose lower frequencies is 
that the decode stage exhibits low GPU utilization, prompting the
governor to reduce the GPU frequency in trying to bring
the utilization to the target range (\S\ref{subse:governors}).
Fig.~\ref{fig:util-freq} (left plot) illustrates the GPU utilization
when the CPU is pinned at each available frequency. Even when the CPU is pinned
to the highest frequency of 2850 MHz, the average GPU utilization remains at
70.9\%, which is below the target range according to Fig.~\ref{fig:Gpu_governor}.
As a result, the GPU governor lowers the GPU frequency.

\textbf{Prefill.}
We repeat the above three sets of experiments for the prefill stage
of LLM inference. 
The results are shown in Fig.~\ref{fig:finding-gpu-prefill}.
We make the following observations.
(1) Fig.~\ref{fig:finding-gpu-prefill}(a) upper
shows unlike decode, the GPU governor achieves close to optimal 
TTFT and energy-per-token for the three LLM models.
This is because 
compared to decode, prefill enjoys higher GPU utilization due to
token batching, which leads the GPU governor to choose sufficiently high
frequencies, as shown in Fig.~\ref{fig:finding-gpu-prefill}(a) lower half.
For instance, the effective GPU frequencies are
680.7, 738.8, and 811.3 MHz for TinyLlama, StableLM, and Llama-2,
respectively; pinning the GPU frequency to the
  optimal frequency 762 MHz for TinyLlama only
reduces the GPU governor's TTFT by 11.2\%, from 3.2 to 2.9
seconds.
(2) Fig.~\ref{fig:finding-gpu-prefill}(b) shows
the GPU governor also achieves close to optimal TTFT and energy efficiency
with various pinned memory frequencies,
again from choosing high frequencies,
\eg the effective GPU frequencies for 2028, 1014, and
421 MHz pinned memory frequencies are 664.8, 682.1, and 733.3 MHz,
respectively.
(3) Fig.~\ref{fig:finding-gpu-prefill}(c) shows the GPU governor can
achieve close to optimal TTFT and energy efficiency by choosing high GPU
frequencies with
various pinned CPU frequencies.

{The reason that the GPU governor selects high frequencies is
  the elevated GPU utilization in the prefill stage, which prompts the GPU
  governor to choose high GPU frequencies.  As shown in
  Fig.~\ref{fig:util-freq} (left plot), when the CPU is pinned to a
  medium frequency of 2188 MHz, the average GPU utilization reaches 82.8\%,
  which is within the target utilization range for the second highest GPU
  frequency of 762 MHz, according to Fig.~\ref{fig:Gpu_governor}.}

\takeaway{ The GPU governor which strives to meet a utilization target range
  tends to operate the GPU at overly low frequencies
  in the decode stage which results in long latency and low
  energy efficiency. In prefill, the GPU utilization is high,
  and the GPU governor can operate the GPU
  at sufficiently high frequencies to achieve near optimal TTFT and
  energy efficiency.
}

\begin{figure*}[tp]
  \hspace{-0.1in}
  \begin{minipage}[c]{.32\linewidth}
\centering
\includegraphics[width=1.1\linewidth, trim=0 15 0 0]{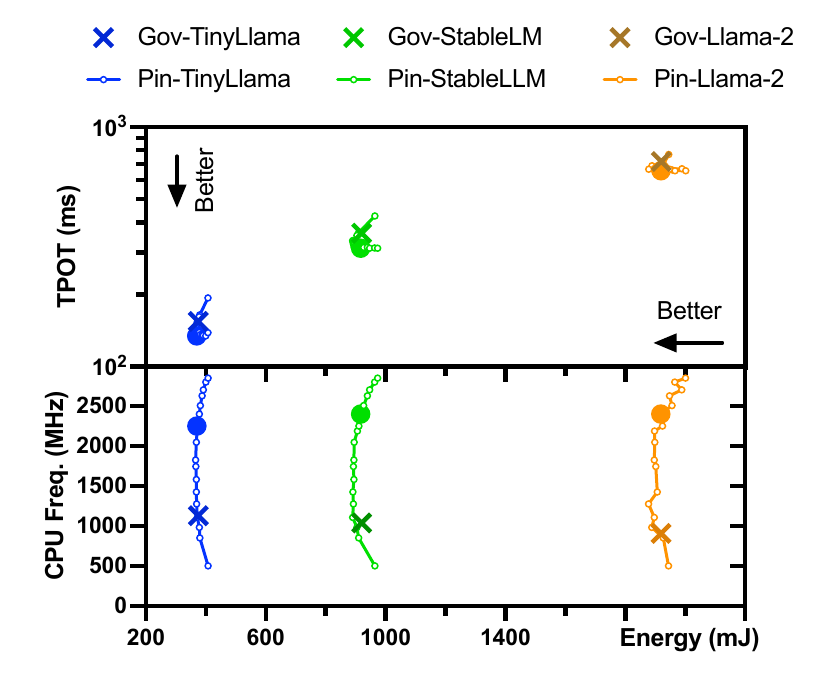}
\subcaption{Various Models}
\label{fig:finding-cpu-decode-a}
\end{minipage}\hfill
\begin{minipage}[c]{.32\linewidth}
\centering
\includegraphics[width=1.1\linewidth, trim=0 15 0 0]{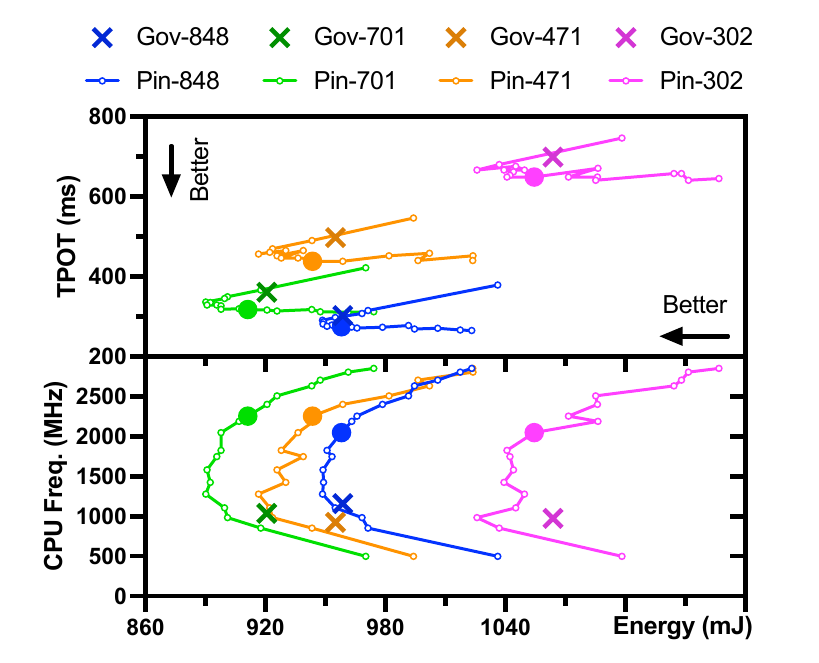}
\subcaption{Varying pinned $f_{GPU}$}
\label{fig:finding-cpu-decode-b}
\end{minipage}\hfill
\begin{minipage}[c]{.32\linewidth}
\centering
\includegraphics[width=1.1\linewidth, trim=0 15 0 0]{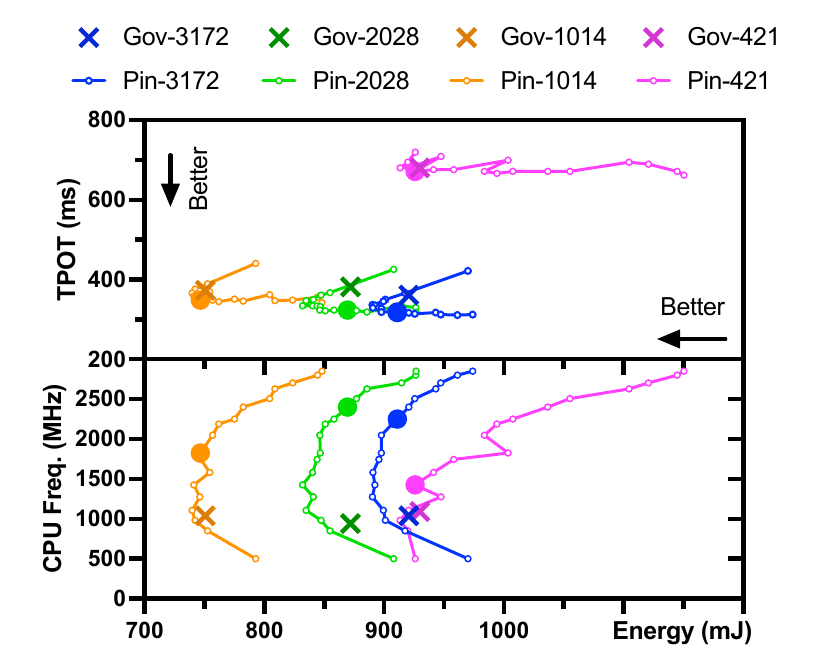}
\subcaption{Varying pinned $f_{MEM}$}
\label{fig:finding-cpu-decode-c}
\end{minipage}
\vspace{-0.15in}
\caption{Decode latency and energy drain of EAS (\texttt{Gov})
  compared with pinning CPU at each available frequency (\texttt{Pin}). We
  set $f_{MEM}=3172$ MHz in (a) and (b), and $f_{GPU}=701$
  MHz in (a) and (c). Plots (b) and (c) are for
  StableLM. {The lowest-latency frequency combinations with the same energy drain as EAS, \texttt{Pin-Opt}, is marked with }\tikzcircle{2pt}.}
\label{fig:finding-cpu-decode}
\vspace{-3mm}
\end{figure*}

\subsection{Memory Governor}
\label{subsec:memgov}

Next, we isolate the impact of the memory governor on LLM inference
from other governors, by pinning CPU and memory frequencies and
compare LLM inference under the default memory governor vs.  when 
pinning the memory to each available frequency using {\tt Pin}.
Fig.~\ref{fig:finding-mem-decode} shows the results for decode.
We make the following observations.
(1) Fig.~\ref{fig:finding-mem-decode}(a) shows 
the results for various models
with pinned GPU and CPU frequencies at 471 MHz and 1826 MHz.
We see the memory governor achieves near optimal inference
latency and energy consumption per token.
For example, for TinyLlama,
it achieves 313.5 mJ energy-per-token and 152.4 ms TPOT,
with an effective frequency of 1019.8 MHz.
Only one pinned memory frequency, at 1539 MHz,
slightly outperforms the memory governor under
the same energy budget; it achieves 309.9 mJ energy per token and 145.1 ms TPOT (4.8\%
lower than the memory governor).
(2)
Similar trends can be observed when we focus on TinyLlama and vary
the pinned GPU frequency (Fig.~\ref{fig:finding-mem-decode}(b)) or pinned CPU frequency
(Fig.~\ref{fig:finding-mem-decode}(c)).
One exception is observed in Fig.~\ref{fig:finding-mem-decode}(b),
when the GPU is pinned at its highest frequency of
848 MHz. The memory governor constantly runs at the highest memory
frequency (3172 MHz) in this case, resulting in high energy
consumption (385.6 mJ per token and 118.2 ms TPOT)
compared to the lowest-energy pinned memory
frequency of 2028 MHz (363.4 mJ per token) with the same TPOT (117.3 ms) as the memory governor.
Similar trends are observed in prefill, and results are omitted due to page limit.

\takeaway{ When fixing the CPU/GPU frequencies,
  the default memory governor can achieve near optimal inference
  latency and energy consumption per token.  }
  
\subsection{EAS is only CPU-Energy Aware}
\label{subsec:eas}

Although the mobile LLM framework offloads most computations to the GPU, as
explained in \S\ref{subsec:inference}, the CPU still plays a key role
during inference and thus its frequency can directly impact inference
performance and energy efficiency. We analyze EAS's impact
by pinning GPU and memory frequencies in the following experiments.

\textbf{Decode.}
Fig.~\ref{fig:finding-cpu-decode} compares inference performance in
the decode stage under EAS with pinning the CPU at each available
frequency.
We observe that EAS consistently achieves higher TPOT and energy consumption
compared to optimal pinned CPU frequency,
regardless of model sizes
(Fig.~\ref{fig:finding-cpu-decode}(a)), pinned GPU frequencies
(Fig.~\ref{fig:finding-cpu-decode}(b)), or pinned memory frequencies
(Fig.~\ref{fig:finding-cpu-decode}(c)).
For instance, when the GPU is pinned at a medium frequency of
471 MHz as shown in Fig.~\ref{fig:finding-cpu-decode}(b), EAS
achieves 955.1 mJ energy per token and 497.5 ms TPOT,
while pinning the CPU at
2252 MHz results in 11.8\% lower TPOT (438.6 ms) with a similar
energy consumption (943.5 mJ). Even pinning the CPU at 1426 MHz results
in 6.5\% lower TPOT with lower energy-per-token (930.2 mJ).

The longer TPOT under EAS can be explaiend by
how it chooses the CPU frequency.
The lower half of above figures show that 
EAS consistently chooses lower CPU frequencies than
the optimal pinned CPU frequency
{that achieves the same energy as EAS.}
For instance, the effective CPU frequencies for TinyLlama, StableLM, and
Llama-2 are 1130.8, 1038.8, and 907.2 MHz,
lower than the optimal pinned CPU frequency of
2252, 2401, and 2401 MHz, respectively.
We also observe that 
CPU frequencies chosen by EAS follow a bimodal distribution
pattern; EAS frequently switches between two values for roughly equal
amount of time during inference: one medium frequency (\eg 1426 MHz)
and one low frequency (\eg 851 or 500 MHz).

\begin{figure*}[!t]
\begin{minipage}[c]{.5\linewidth}
\centering
\includegraphics[width=\linewidth, trim=0 10 0 0]{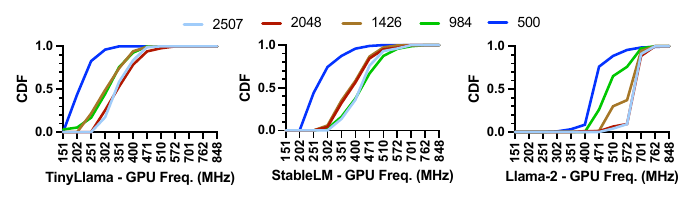}
\subcaption{CPU frequency affects GPU governor}
\label{fig:motiv-p4-pin-one-side-a}
\end{minipage}\hfill
\begin{minipage}[c]{.5\linewidth}
\centering
\includegraphics[width=\linewidth, trim=0 15 0 0]{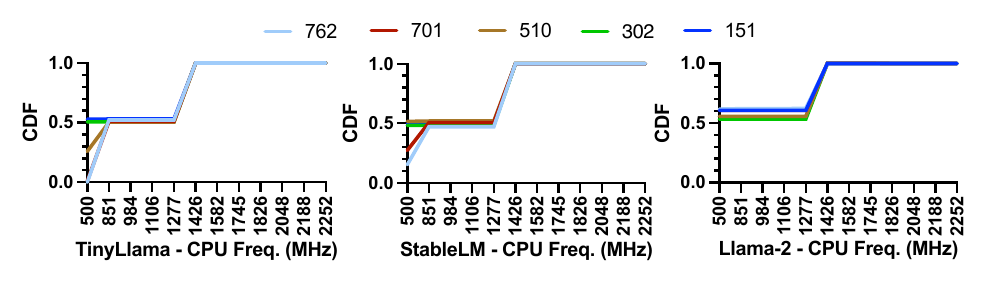}
\subcaption{GPU frequency affects EAS}
\label{fig:motiv-p4-pin-one-side-c}
\end{minipage}
\vspace{-3mm}
\caption{CDF of the amount of time that GPU (left) or CPU (right)
runs at a certain frequency under GPU governor or EAS, respectively. Results are collected in the decode phase with three models served by llama.cpp. We set $f_{MEM}$=3172 MHz.}
\label{fig:motiv-p4-pin-one-side}
\vspace{-3mm}
\end{figure*}

\textbf{Prefill.}
{For the prefill stage, similar to the decode stage, EAS consistently achieves higher TTFT (for
the same energy budget) or higher energy consumption (for the same TTFT),
regardless of model size,
pinned GPU frequency, or
pinned memory frequency,
again due to consistently choosing low frequencies in all settings.
Due to page limit, detailed results of prefill latency and energy drain under
EAS compared with pinning the CPU at each available frequency are shown in
Fig.~\ref{fig:finding-cpu-prefill} in Appendix~\hyperref[appendix:eas]{A}.}

{The reason that EAS governor chooses overly low
  frequencies is the low CPU utilization in either
  the prefill or decode stage.  As shown in Fig.~\ref{fig:util-freq}
  (right plot), even when the GPU is pinned to the highest frequency of 848
  MHz, the average CPU utilizations for the prefill and decode stages
  under a pinned CPU frequency of 2188 MHz are 17.5\% and 25.7\%, causing
  the EAS governor to lower the CPU frequency.}

\takeaway{ EAS operates the CPU at overly low frequencies in both
  decode and prefill stages of LLM inference which degrades inference
  latency and energy efficiency.
}

\begin{figure}[!t]
\centering
\includegraphics[width=\linewidth, trim=0 10 0 0]{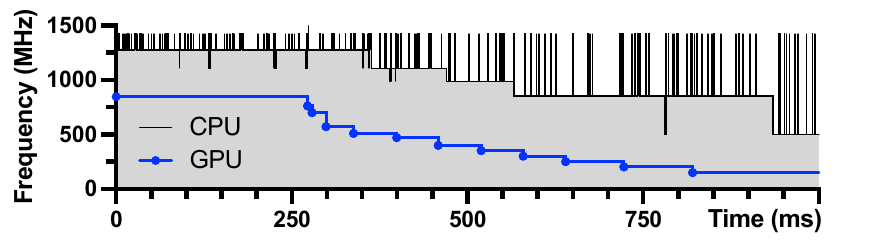}
\caption{Runtime trace of CPU and GPU frequencies showing the antagonistic effect.
\cut{GPU frequency is pinned at 848 MHz at the beginning, then unpinned after 250 milliseconds. The CPU is controlled by the EAS governor.}}
\label{fig:motiv-staircase}
\vspace{-6mm}
\end{figure}

\subsection{Antagonistic EAS and GPU Governors}
\label{subsec:eas-gpu}

From \S\ref{subsec:gpu-gov} and \S\ref{subsec:eas}, we learned that
both the GPU governor and EAS, when acting alone, tend to choose
overly low frequencies, which result in higher inference latency (for
the same energy budget) than the optimal pinned frequencies.
These findings in turn raise an important question: 
when both governors operate concurrently, do they 
{\em antagonistically} affect each other, \ie
cascadingly driving both GPU and CPU frequencies lower in
a downward spiral. Such bevahior could severely degrade
inference performance and energy efficiency.
We design controlled experiments to answer this question.

\textbf{Lower CPU frequency leads to lower GPU frequency.}
We first pin the CPU frequency at different values
and observe the trend of GPU frequencies chosen
by the GPU governor during decode of various models.
Figures~\ref{fig:motiv-p4-pin-one-side}(a) 
show the distribution of time the GPU is at different frequencies 
during inference for each pinned CPU frequency.
We see that 
as we lower the  pinned CPU frequency from
2507 MHz to 500 MHz, the CDF curve of GPU frequencies
clearly moves towards left, meaning that the GPU governor spends longer time at
lower frequencies.
For instance, while decoding TinyLlama, the most
frequently chosen frequency by the GPU governor is
351 MHz (41\% of the time) when the CPU is pinned at 2507 MHz,
but is lowered to
202 MHz (44\% of the total time) when the CPU is pinned at 500 MHz.
Second, the trend is consistent across 
different models (Fig.~\ref{fig:motiv-p4-pin-one-side}(a)).
In decoding Llama-2, the GPU governor mostly operates
at 701 MHz (82\% of time) when the CPU is pinned at 2507 MHz,
but at
471 MHz (67.6\% of time) when the CPU is pinned at 500 MHz.

\textbf{Lower GPU frequency leads to lower CPU frequency.}
Conversely,
Figures \ref{fig:motiv-p4-pin-one-side}(b) 
show that lowering the pinned GPU frequency leads to EAS lowering 
the CPU frequencies chosen.
As mentioned in \S\ref{subsec:eas}, the CPU
frequency controlled by EAS usually fluctuates between two
frequencies for roughly equal amount of time, resulting in two flat
regions in the CDF curves in the figures.
For example, with TinyLlama, EAS mostly chose 
851 MHz CPU frequency (52.2\% of the time) when the GPU is pinned to 762 MHz,
but
500 MHz (52.9\% of the time) when the GPU is pinned at 151 MHz.
Further, this 
relationship
is consistent across different models (Fig.~\ref{fig:motiv-p4-pin-one-side}(b)).

\textbf{The antagonistic effect.}  We further capture the antagonistic effect
between EAS and GPU governors during LLM inference
in real time and visualize it
in Fig.~\ref{fig:motiv-staircase}.  The GPU
is pinned to the highest frequency of 848 MHz at the beginning,
then unpinned after 250 ms (\ie let the default GPU governor control
GPU frequency). We let the default EAS
governor control CPU frequencies throughout the experiment. 
As illustrated in
Fig.~\ref{fig:motiv-staircase}, immediately after the GPU is
unpinned, the GPU governor drops its  frequency from 848 to 510 MHz
{in 4 steps between 250 and 338 ms}. 
During this period, the CPU
frequency is stabilized at 1277 MHz.  At 363 ms, the CPU frequency drops
from 1277 to 1106 MHz, which in turn drives the GPU governor to lower
the GPU frequency from 510 to 471 MHz at 399 ms.  The antagonistic
effect continues, ultimately driving the GPU governor to lower the GPU frequency
to its minimum of 151 MHz at 821 ms, and the CPU governor to lower its frequency
to its minimum of 500 MHz at 935 ms.
{Due to page limit, the illustration of the antagonistic
  effect where the default GPU governor control the GPU frequency while
  the CPU frequency is initially pinned and then unpinned is shown in
  Fig.~\ref{fig:motiv-staircase-cpu} in Appendix~\hyperref[appendix:spiral]{B}.}

\textbf{The root cause.}
The root cause of the antagonistic effect lies in the independent
frequency scaling of each governor, as they attempt to meet their
respective utilization targets.
For instance, the GPU governor
dynamically adjusts the GPU frequency to align GPU utilization
with the vendor-defined target range (Fig.~\ref{fig:Gpu_governor} in \S\ref{subse:governors}).
The antagonistic effect begins with low
utilization of either component while running the inference engine. Suppose the CPU utilization is low, in
response, the CPU governor lowers the CPU frequency to increase
CPU utilization.  However, the lower CPU frequency slows
down the OpenCL runtime running on the CPU (\S\ref{subsec:inference}), delaying issuing GPU
tasks and hence reducing the GPU utilization (see next paragraph).  To compensate,
the GPU governor lowers the GPU frequency,
further extending the waiting time between GPU task executions. This
prolonged delay reduces CPU utilization, prompting the CPU governor to
lower the CPU frequency even further, perpetuating the cycle.

To illustrate that lowering the frequency of one component lowers the
utilization of the other component, Fig.~\ref{fig:util-freq} shows the
average CPU utilization when the GPU frequency is pinned at different
levels, and GPU utilization when the CPU frequency is pinned at
different levels.  Results are collected while
inferencing TinyLlama model on Pixel 7. In the decode stage, as CPU frequency decreases
from 2850 MHz to 500 MHz, the average GPU utilization drops from 70.9\% to 52.9\%. Similarly, as GPU frequency decreases from 848
MHz to 151 MHz, the average CPU utilization drops from 25.7\% to 7.9\%.

\takeaway{ EAS and the GPU governors can trigger a ``downward spiral'' by
  cascadingly driving each other to choose lower CPU/GPU frequencies.
  Avoiding such antagonistic effect between independently acting governors
  requires a holistic energy-efficient governor for managing both the GPU and CPU.
}

\vspace{0.1in}
\section{FUSE: a Unified Energy-aware Governor}
\label{sec:design}

Motivated by the limitations of independent governors shown in
\S\ref{sec:interplay}, we design \name, a unified energy-aware governor
for optimizing the energy-efficiency of LLM inference on mobile devices.
Given an LLM model, the goal of \name is to find and configure 
CPU/GPU/memory to run at the frequency combination that
\textbf{(G1)} minimizes the inference latency given an energy budget\footnote{We assume
  the energy budget is input by the user, who could specify it in
  absolute terms, \eg 40\% battery level drop in 4 hours, or abstract
  terms such as low/medium/high.},
or \textbf{(G2)} minimizes the energy consumption given an inference latency
target, \ie TTFT for prefill and TPOT for decode.

{\bf Design overview.}
{
We observe that
in LLM-powered personal services on mobile devices, the same LLM model (\eg embedded
in an APP or the OS~\cite{foundationmodel:mobicom2024})
is typically used over an extended period of time.
This motivates a simple, offline-profiling-based approach.
During APP or OS installation, \name efficiently searches for optimal
frequency configurations for the prefill and decode stages. These
configurations are then applied to every model inference at runtime,
triggered by notifications from the inference framework indicating the
start and end of these phases.

For efficient frequency search, we observe that the optimal frequency
configuration for a given LLM model is input-content-agnostic and primarily
affected by the prefill length.  Based on this,
\name categorizes prefill lengths into five distinct ranges, and
performs frequency searches for one sampled decoding length and five
representative prefill lengths-—one from each range.
Below we detail
the frequency search process for one setting.
}

\subsection{Efficient Frequency Search}

The design of \name's frequency search is motivated by the findings in \S\ref{sec:interplay} that
(1) the default CPU/GPU governors tend to cascadingly drive each other's frequency down,
(2)
among GPU/CPU/memory frequencies, GPU frequency is the dominant factor
affecting inference latency and energy efficiency
for primarily GPU-based 
LLM inference.
These observations motivate a two-step search process for
optimal frequency combinations: 
(1) \name mitigates the antagonistic effect by first searching for optimal
GPU frequencies, by pinning the GPU at candidate frequencies;
(2) It fine-tunes the CPU frequency by exploring CPU frequencies
while pinning the GPU at the selected GPU frequencies (at most two) from step 1.
We leave the memory governor to
its default settings as our findings in \S\ref{subsec:memgov}
indicate that it can achieve near-optimal inference latency and energy
efficiency.

\begin{figure}[!t]
\centering
\begin{minipage}[c]{.48\linewidth}
\centering
\includegraphics[width=\linewidth, trim=0 10 0 0]{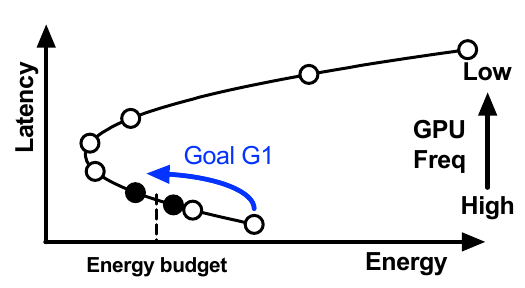}
\subcaption{Goal \textbf{G1}}
\label{fig:search-goal-g1}
\end{minipage}\hfill
\begin{minipage}[c]{.48\linewidth}
\centering
\includegraphics[width=\linewidth, trim=0 10 0 0]{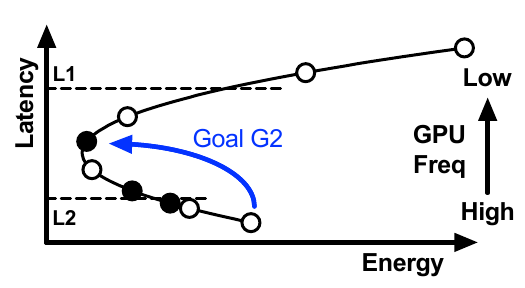}
\subcaption{Goal \textbf{G2}}
\label{fig:search-goal-g2}
\end{minipage}
\vspace{-3mm}
\caption{Selected GPU frequencies (\tikzcircle{2pt} solid points) in Step 1.}
\label{fig:search-goal}
\vspace{-6mm}
\end{figure}

\begin{figure*}[!t]
\begin{minipage}[c]{\linewidth}
\centering
\includegraphics[width=0.98\linewidth, trim=0 8 0 0]{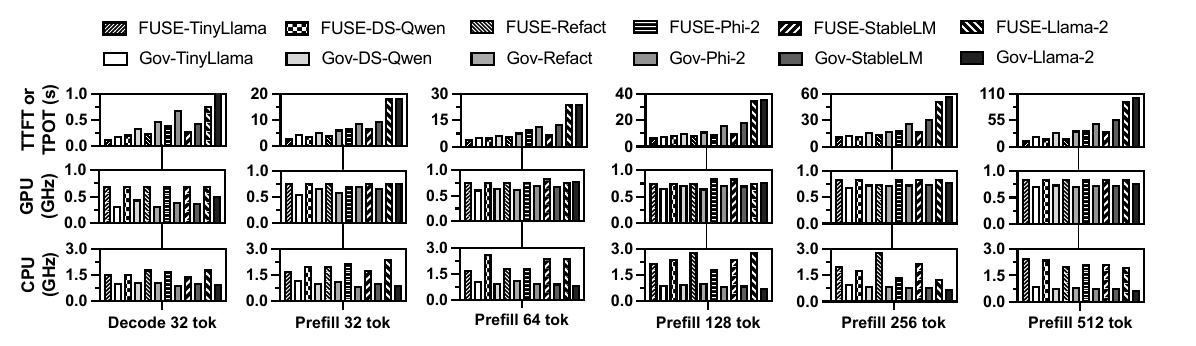}
\end{minipage}
\vspace{-3mm}
\caption{Performance comparison of \name with \texttt{Gov} (the default governors),
for goal \textbf{G1}.
Energy-per-token budget is the energy draw under the default governors.
}
\label{fig:fuse-faster}
\vspace{-3mm}
\end{figure*}

\textbf{Step 1: GPU frequency search.}
To minimize inference latency given an energy budget (\ie goal \textbf{G1}),
since fixing CPU/memory frequencies,
changing the GPU frequency results in a U-shape energy-per-token curve
(as shown in Fig.~\ref{fig:finding-gpu-decode}),
our search begins from the highest GPU frequency and decrements it one
step at a time.
The search stops at
the first GPU frequency that achieves a lower energy-per-token than
the energy budget, as the following even lower GPU frequencies
will lead to higher inference latency even if they can meet the energy budget,
as shown in Fig.~\ref{fig:search-goal-g1}.

For a given GPU frequency, Fig.~\ref{fig:finding-cpu-decode}
and \ref{fig:finding-cpu-prefill} showed that changing
the CPU
frequency results in a U-shape energy-per-token curve.
Thus for \textbf{G1}, in step 1 \name takes both the first GPU frequency $F$
whose energy is within the energy budget and the previous GPU frequency
$F'$ whose energy is above the energy budget,
as there may exist CPU frequencies for $F'$ that achieve a total energy
within the energy budget.

To minimize energy consumption given an inference latency target (\ie
goal \textbf{G2}), the search first finds the minimum-energy
frequency, defined as the GPU frequency with the lowest energy draw
with no latency constraint.  The search starts from the highest GPU
frequency and stops at the GPU frequency that draws more energy than
the previous frequency, \ie the minimum-energy frequency,
as shown in Fig.~\ref{fig:search-goal-g2}.
Next, if the latency target is higher than the latency at the minimum-energy
frequency, \eg L1 in Fig.~\ref{fig:search-goal-g2}, \name chooses the
minimum-energy frequency; otherwise, it chooses two consecutive
GPU frequencies whose latencies are higher and lower than the latency
target, \eg L2 in Fig.~\ref{fig:search-goal-g2}.

\textbf{Step 2: CPU frequency search.}
Since the default CPU governor tends to run at overly-low frequencies
(\S\ref{subsec:eas}), for \textbf{G1}, in the second step, \name
searches for the optimal CPU frequency while pinning the GPU 
at each (at most two) candidate GPU frequency chosen in Step 1.
The search starts from the highest CPU frequency and stops at the first CPU frequency
that achieves a lower energy-per-token than the energy budget.
It then outputs the CPU/GPU frequency combination that achieves the
lowest inference latency.
Similarly, for \textbf{G2}, the search starts from the highest CPU frequency and stops at the first 
CPU frequency that achieves a higher
latency than the latency target, and outputs the CPU/GPU frequency
combination that achieves the lowest energy-per-token.

\begin{figure}[!t]
\captionof{table}{Evaluated decoder models in llama.cpp.} 
\vspace{-3mm}
\begin{minipage}[c]{\linewidth}
\centering
\resizebox{\textwidth}{!}{
\begin{tabular}{ *{5}{c} }
    \toprule
    \bf Model & \bf \#Layers & \bf Hidden size & \bf Size & \bf Device \\
    \midrule
    TinyLlama \cite{tinyllama} & 22 & 2048 & 1.1B & Pixel 7 \\
    DeepSeek-R1-Distill-Qwen \cite{deepseekai2025deepseekr1incentivizingreasoningcapability} & 28 & 1536 & 1.5B & Pixel 7 \\
    Smallcloudai Refact-fim & 32 & 2048 & 1.6B & Pixel 7 \\
    StableLM-Zephyr & 32 & 2560 & 2.7B & Pixel 7 \\
    Microsoft Phi-2 & 32 & 2560 & 2.7B & Pixel 7 pro \\
    Meta Llama-2 \cite{touvron2023llama2openfoundation} & 32 & 4096 & 6.7B & Pixel 7 pro \\
    \bottomrule
\end{tabular}}
\end{minipage}
\label{tab:profile-model}
\vspace{-4mm}
\end{figure}

\subsection{Evaluation Results}

We prototyped \name on Android to support the \LLAMACPP~\cite{llamacpp} framework (version: tag \texttt{b2202}) in 2K lines of Python code.
The same
platform described in \S\ref{sec:methodology} is used to evaluate the
performance of \name.
We use the energy drain and inference latency under the default governors
as the
energy budget and latency target.
We evaluate \name with a set of popular LLM models
in 4-bit quantization
as listed in
Table~\ref{tab:profile-model}.

\textbf{Dataset and baseline.}
We randomly sample 200 requests from the ShareGPT dataset
with prefill length no larger than 512
tokens
and decode length no larger than 256
tokens (to fit the memory size of the test devices).
The average prompt length and decode length of our sampled dataset are
232.4 and 70.0
tokens, respectively.
The performance of \name is compared with that of 
the default governors, denoted as \texttt{Gov}.

\begin{figure}[tp]
\begin{minipage}[c]{\linewidth}
\centering
\includegraphics[width=1.05\linewidth, trim=10 8 10 0]{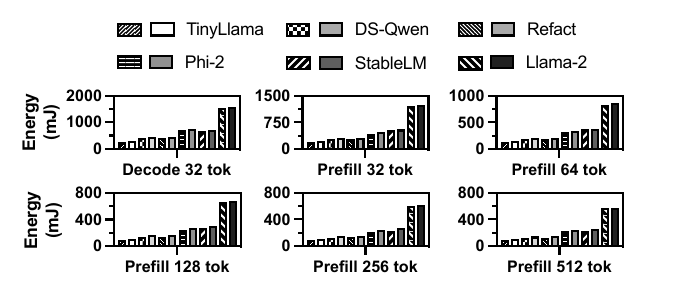}
\end{minipage}
\vspace{-2mm}
\caption{Energy-per-token comparison of \name with \texttt{Gov} (the default governors),
for goal \textbf{G2}.
Latency target is the TTFT or TPOT under the default governors.
}
\label{fig:fuse-efficient}
\vspace{-6mm}
\end{figure}

\textbf{Effectiveness of frequency search.}
We first evaluate the effectiveness of frequency search
for the six settings 
(\ie decode with 32 tokens and prefill with \{32, 64, 128, 256, 512\} tokens).
Fig.~\ref{fig:fuse-faster} compares \name's inference latency against that of \texttt{Gov} for goal \textbf{G1}.
We see that while inferencing with the same energy-per-token
as \texttt{Gov}, \name reduces TPOT and TTFT by 41.0\% and 24.8\%
averaged across all models by running the CPU and GPU at the optimal
frequency combination.
For instance,
while decoding DeepSeek-R1-Distill-Qwen (shortened as DS-Qwen) with the same energy-per-token (460.5 mJ with \name and 459.0 mJ with \texttt{Gov}), \name reduces TPOT by 33.8\% (from 346.8 ms to 229.6 ms) by setting the GPU frequency at 701 MHz (compared to 448.5 MHz with \texttt{Gov}) and the CPU frequency at 1582 MHz (compared to 1134.5 MHz with \texttt{Gov}).
Fig.~\ref{fig:fuse-efficient} compares \name's inference energy-per-token against
that of \texttt{Gov} for goal \textbf{G2}. We observe that while
prefilling with the same TTFT as \texttt{Gov} or decoding with TPOT no higher than the
that of \texttt{Gov}, \name reduces energy-per-token by 6.9\% and
10.3\% averaged across all models in the prefill and decode stage,
respectively.
{Due to page limit, the CPU/GPU frequencies found by \name for goal \textbf{G2} are
shown in Fig.~\ref{fig:fuse-efficient-freq} in Appendix~\hyperref[appendix:g2]{C}.}

\begin{figure}[tp]
\begin{minipage}[c]{0.95\linewidth}
\centering
\includegraphics[width=\linewidth, trim=0 8 0 0]{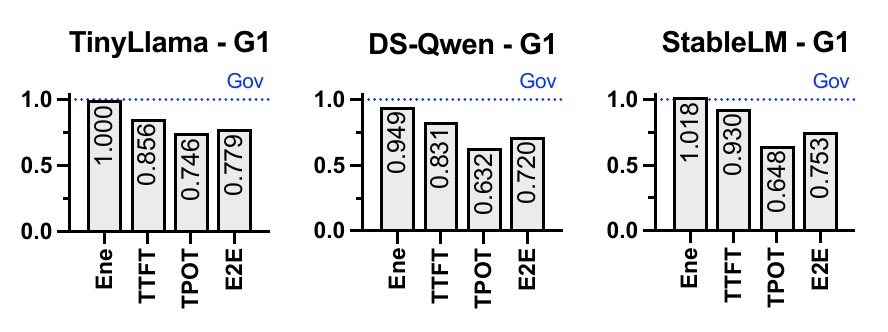}
\end{minipage}
\newline
\begin{minipage}[c]{0.95\linewidth}
\centering
\includegraphics[width=\linewidth, trim=0 8 0 0]{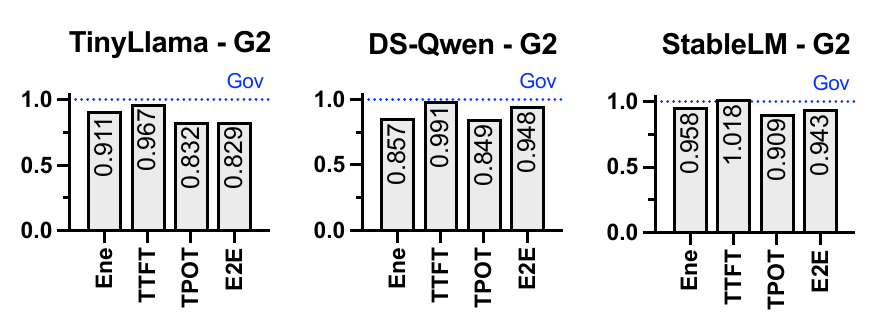}
\end{minipage}
\vspace{-4mm}
\caption{\name's energy consumption and performance normalized to default governors for
goals \textbf{G1} and \textbf{G2} {on the ShareGPT trace}.
}
\label{fig:end2end-g1g2}
\vspace{-6mm}
\end{figure}

\textbf{Performance on real trace.}
For goal \textbf{G1}, Fig.~\ref{fig:end2end-g1g2} top row compares \name's energy
consumption and latency normalized against that of \texttt{Gov} in
serving the 200 sampled inference requests from ShareGPT.  For TinyLlama, while
consuming the same amount of total energy (738.1 mAh with \texttt{Gov}
and 737.8 mAh with \name), \name reduces average TTFT, TPOT, and E2E
latency by 14.4\% (from 10.56 to 9.04 s), 25.4\% (from 210.7 to
157.2 ms), and 22.1\% (from 25.2 to 19.6 s), respectively. Similarly, for
DeepSeek-R1-Distill-Qwen, while consuming the
same amount of total energy (1164.4 mAh with \texttt{Gov} and 1104.7
mAh with \name), \name reduces TTFT, TPOT and E2E latency by 16.9\%,
36.8\%, and 28.0\%, respectively.  For the larger 2.7B StableLM model,
while consuming the same amount of total energy,
\name reduces TTFT, TPOT and
E2E latency by 7.0\% (from 24.2 to 22.5 s), 35.2\% (from 492.5
to 319.2 ms) and 24.7\% (from 58.5 to 44.0 s), respectively.

For goal \textbf{G2}, 
for TinyLlama, while inferencing at
the same target average TTFT (10.56 s with \texttt{Gov} and 10.22
s with \name),
\name reduces the total energy draw by 8.9\% (from 738.1 to 672.3 mAh).
Note that \name's TPOT and E2E latency results are both lower
than \texttt{Gov}, by 16.8\% and 17.1\% respectively.
For DS-Qwen and StableLM models, while inferencing at the
same average TTFT with lower TPOT and E2E latency, \name reduces the
energy draw by 14.3\% (from 1164.4 to 997.76 mAh) and 4.2\% (from 1772.3 to 1698.6
mAh), respectively.

\textbf{Search cost.}
For each model,
\name performs profiling-based frequency search for each of the six settings
for either goal \textbf{G1} or \textbf{G2}.
{For goal {\bf G1},
it only performs on average 2.4 and 5.1 inferences per setting (\ie 14.5
and 30.8 inferences in total per model)} across the 6 models
in Step 1 and Step 2, respectively---a reduction
of 374x from the 2808 total CPU/GPU/memory frequency combinations.
{
Multiplied by per-inference time, which differs
across the models and settings, ranging from 23.4 to 104.0 seconds,
frequency search finishes in 17.7, 43.1, and 78.5 minutes}
for all settings for TinyLlama, StableLM, and Llama-2 models.
For goal \textbf{G2}, \name takes on average {3.6 and 4.8 inferences per setting (\ie 21.8 and 28.8 inferences in total per model)} in
Step 1 and Step 2, averaged across the 6 models.
It spends more steps in Step 1 than \textbf{G1},
in finding the minimum-energy GPU frequency.
{Multiplied by per-inference time, frequency search finishes in 19.7, 48.1, and 87.7 minutes} for all settings for the models.

\section{Related Work}
\label{sec:related_work}

\textbf{Mobile LLM profiling and benchmarking.}
Laskaridis \etal~\cite{meltingpoint} performed the first systematic on-device
LLM performance and energy efficiency profiling. Li \etal~\cite{edgefm-bench}
focused on profiling the accuracy, latency, and memory footprint of
mobile LLM inferencing without power measurement. The benchmark from Xiao
\etal~\cite{xiao2024largelanguagemodelperformance} covers many
mobile devices and perspectives, including the impact of CPU scheduling on LLM
inference performance. However, none of these works analyzed the impact of DVFS
governors on LLM performance and energy efficiency.

\textbf{Mobile LLM performance optimization.}
A number of works explored optimizing LLM inference on mobile devices.
Firstly, several works proposed mobile-friendly LLM model designs or adaptations.
These include optimizing the memory footprint of the LLM model via
smaller LLM model architecture~\cite{mobilellm,minicpm-v}, model weight
quantization~\cite{mobilequant}, and model reuse across different
tasks~\cite{llmfirmware}. On the other hand, the high memory footprint of LLM
inference is tackled by designing mobile LLM inference frameworks that load
model weights on-demand~\cite{powerinfer2,llmcad,edgemoe,sti}. Finally, special
hardware, \eg NPU, on mobile devices are utilized to improve the LLM inference
performance~\cite{mllm-npu}. However none of the work study the energy
consumption of LLM inference on mobile devices.

\textbf{Mobile DVFS optimizations.}
Several prior works optimized DVFS for different scenarios, \eg avoiding
thermal throttling~\cite{ztt,maestro,cartad}, adapting to concurrent
tasks~\cite{GearDVFS}, and edge computing~\cite{DRLDO}. On the other hand, DVFS
optimizations have been proposed for specific applications, \eg DNN
inference~\cite{dvfo,10261988,park2017ml}, where optimal frequency combinations are
searched. However, none of the previous work (including \cite{he2018deadline,park2015cooperative,hsieh2015memory}) have examined the intricate
interplay among DVFS governors
in mobile OSes
and its impact on LLM inference performance and
energy efficiency.

\section{Conclusions}
\label{sec:conclusions}

In this paper, we presented to our knowledge the first in-depth study
of the interplay of mobile CPU, GPU, and memory governors during LLM
inference.
Our study shows that the triplet governors used in mobile OSes such as Android
can result in 23\% to 40.4\% longer prefilling and decoding latency compared to
optimal combinations of CPU/GPU/memory frequencies under the same energy budget,
or 5.0\% to 16.6\% more energy consumption under the same latency.
Via controlled experiments, we further uncovered the root causes 
as (1)
acting alone,
these governors tend to choose lower frequencies,
(2) acting concurrently, they can trigger a "downward spiral" of the CPU/GPU frequencies.
Finally,
we presented a unified energy-aware governor, \name, that
is shown to reduce TTFT and TPOT of LLM inference
by 7.0\%-16.9\% and 25.4\%-36.8\%
on average
for various mobile LLM models.

% \balance

\bibliographystyle{ACM-Reference-Format}
\bibliography{references}

%\newpage
%\appendix

% \newpage
\appendix

\begin{figure*}[!t]
  \hspace{-0.1in}
  \begin{minipage}[c]{.32\linewidth}
\centering
\includegraphics[width=1.0\linewidth, trim=0 15 0 0]{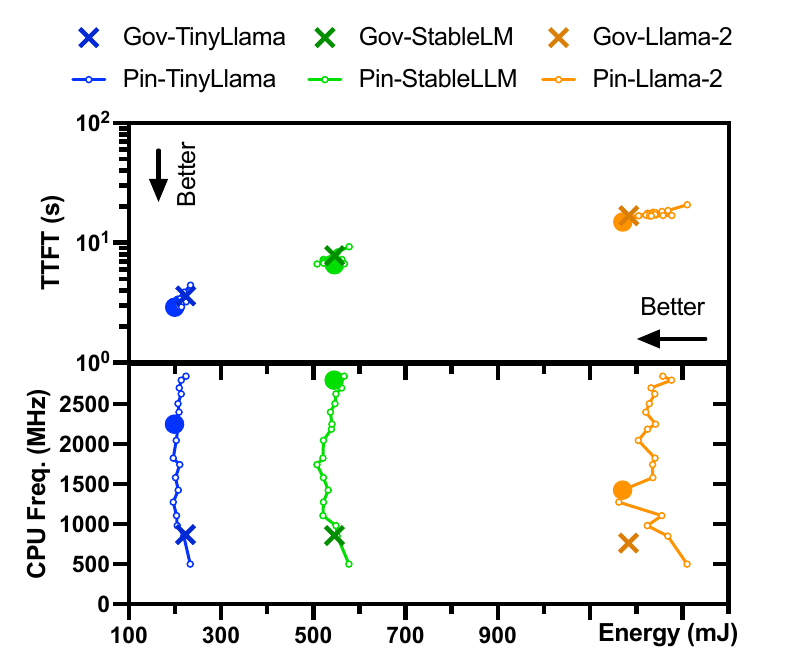}
\subcaption{Various Models}
\label{fig:finding-cpu-prefill-a}
\end{minipage}\hfill
\begin{minipage}[c]{.32\linewidth}
\centering
\includegraphics[width=1.0\linewidth, trim=0 15 0 0]{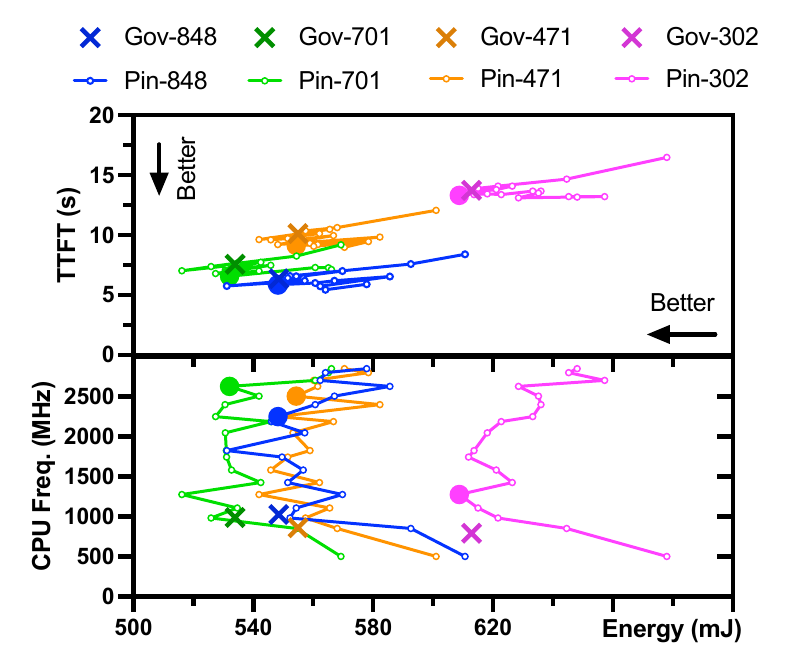}
\subcaption{Varying pinned $f_{GPU}$}
\label{fig:finding-cpu-prefill-b}
\end{minipage}\hfill
\begin{minipage}[c]{.32\linewidth}
\centering
\includegraphics[width=1.0\linewidth, trim=0 15 0 0]{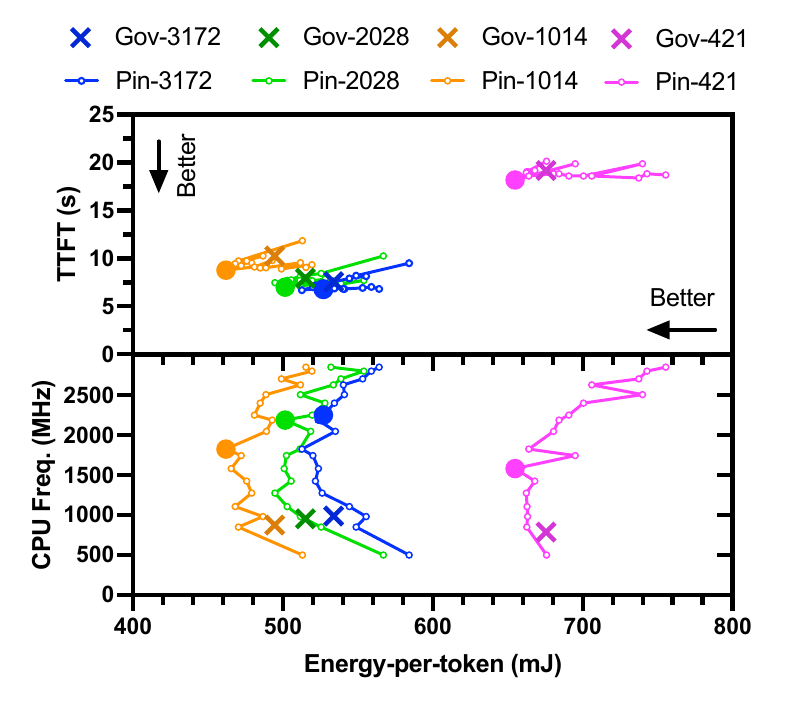}
\subcaption{Varying pinned $f_{MEM}$}
\label{fig:finding-cpu-prefill-c}
\end{minipage}
\vspace{-0.15in}
\caption{Prefill latency and energy drain of EAS
  (\texttt{Gov})
  compared with pinning the CPU at each available frequency (\texttt{Pin}). We
  set $f_{MEM}=3172$ MHz in (a) and (b), and set $f_{GPU}=701$
  MHz in (a) and (c). Results in (b) and (c) are for
  StableLM. {\texttt{Pin-Opt} is marked with }"\tikzcircle{2pt}".}
\label{fig:finding-cpu-prefill}
% \vspace{-4mm}
\end{figure*}

\section*{Appendix}

\subsection*{A: Additional EAS Governor Results}
\label{appendix:eas}

In \S\ref{subsec:eas}, 
we analyzed EAS's impact on LLM inference 
by pinning GPU and memory frequencies for the decode stage.
Here we show the results for the prefill stage.

\textbf{Prefill.}
For the prefill
stage, similar to the decode stage, EAS consistently achieves higher TTFT (for
the same energy budget) or energy consumption (for the same TTFT),
regardless of model sizes (Fig.~\ref{fig:finding-cpu-prefill}(a)),
pinned GPU frequencies (Fig.~\ref{fig:finding-cpu-prefill}(b), or
pinned memory frequencies (Fig.~\ref{fig:finding-cpu-prefill}(c)), due to consistently choosing low frequencies in all settings, as
shown in the lower half of Fig.~\ref{fig:finding-cpu-prefill}. For instance, the TTFT
under EAS for TinyLlama, StableLM, and Llama-2 are 3.6, 7.8, and 16.8
seconds, which can be reduced by 18.9\%, 16.2\%, and 11.2\% by pinning
CPU to the optimal frequencies of 2802, 2802, and 1426 MHz with a similar
energy consumption, respectively. The longer TTFT under EAS can be
explained by the fact that EAS chooses overly low CPU
frequencies. Specifically, the effective CPU frequencies for three
models are 870.7, 858.2, and 763.4 MHz.

\vspace{0.1in}
\subsection*{B: Additional Antagonistic Effect Results}
\label{appendix:spiral}

\begin{figure}[!h]
\centering
\includegraphics[width=\linewidth, trim=0 10 0 0]{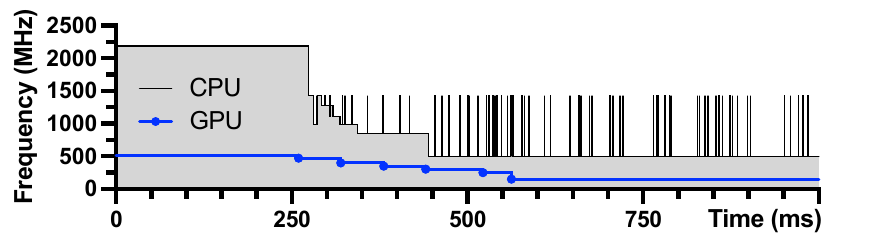}
\caption{Runtime trace of CPU and GPU frequencies showing the antagonistic effect.
}
\label{fig:motiv-staircase-cpu}
\end{figure}

In \S\ref{subsec:eas-gpu}, 
we demonstrated the antagonistic effect between the EAS and GPU governors in real time
by pinning and unpinning the GPU frequency.

To further illustrate the antagonistic effect between the EAS and GPU
governors, we let the default GPU governor control GPU frequencies
throughout the experiment and pin the CPU to 2188 MHz at the beginning,
then unpin it at 250 ms, \ie let the default EAS governor control CPU
frequencies.  As illustrated in Fig.~\ref{fig:motiv-staircase-cpu},
immediately after the CPU is unpinned, the CPU governor drops its
frequency from 2188 to 984 MHz at 287 ms.  During this period, the GPU
frequency is at 471 MHz.  At 320 ms, the GPU frequency drops from 471
to 400 MHz, which in turn drives the CPU governor to lower the CPU
frequency from 984 to 851 MHz at 343 ms.  The antagonistic effect
continues, ultimately driving the CPU governor to lower its frequency
to its minimum of 500 MHz at 445 ms, and the GPU governor to lower its
frequency to its minimum of 151 MHz at 563 ms.

\vspace{0.1in}
\subsection*{C: Additional Evaluation Results for Goal \textbf{G2}}
\label{appendix:g2}

\begin{figure}[h]
\begin{minipage}[c]{\linewidth}
\centering
\includegraphics[width=\linewidth, trim=10 8 10 0]{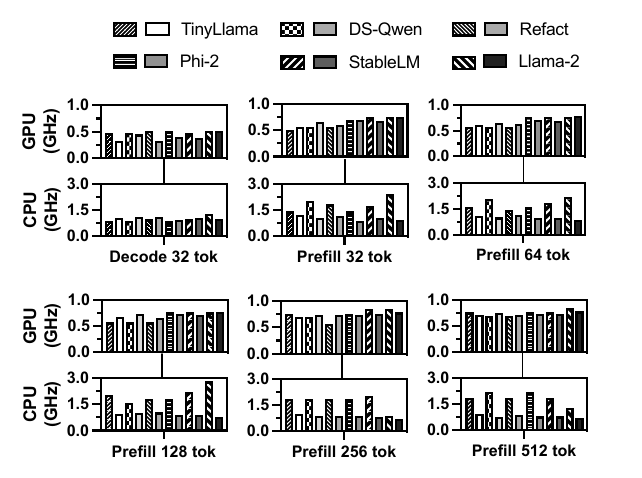}
\end{minipage}
\caption{CPU/GPU frequency comparison of \name with \texttt{Gov} (the default governors),
for goal \textbf{G2}.
Latency target is the TTFT or TPOT under the default governors.
}
\label{fig:fuse-efficient-freq}
\end{figure}

Fig.~\ref{fig:fuse-efficient-freq}
shows the CPU/GPU frequencies
that correspond to the energy-per-token results previously presented in
Fig.~\ref{fig:fuse-efficient}.  We see that while inferencing with the
same TTFT as \texttt{Gov}, \name reduces the energy-per-token by
searching and setting the CPU/GPU to the optimal frequency
combination.  For example, while prefilling 32 tokens with
DeepSeek-R1-Distill-Qwen (shortened as DS-Qwen), by setting the GPU
and CPU frequencies to 572 and 2048 MHz (compared to 663.4 and 1045.6
MHz with \texttt{Gov}), \name reduces the energy-per-token by 10.3\%
(from 310.3 to 278.2 mJ) while inferencing at the same or lower TTFT (5.6 s
with \texttt{Gov} and 5.4 s with \name).  While decoding 32 tokens
with DS-Qwen, by setting the GPU and CPU frequencies to 471 and 851
MHz (compared to 448.5 and 1134.5 MHz with \texttt{Gov}), \name
reduces the energy-per-token by 7.8\% (from 459.0 to 423.3 mJ) while
inferencing at the same or lower TPOT (346.8 ms with \texttt{Gov} and 298.4 ms
with \name).

\end{document}